\documentclass[aps, prd, superscriptaddress, preprintnumbers]{revtex4}

\usepackage{graphicx,amsfonts,amsmath,amssymb,amstext}
\usepackage{float,wrapfig}
\usepackage{subfigure, psfrag}
\usepackage{dsfont}
\usepackage{color}
\usepackage{bm}

\newcommand{\be}{\begin{equation}}
\newcommand{\ee}{\end{equation}}
\newcommand{\ba}{\begin{eqnarray}}
\newcommand{\ea}{\end{eqnarray}}
\newcommand{\sign}[1]{\,\mbox{sgn}\left({#1}\right)}

\definecolor{purple}{rgb}{0.8,0,0.6}
\definecolor{darkgreen}{rgb}{0.00,0.6,0.00}

\begin{document}

\title{Anomalous transport properties of Dirac and Weyl semimetals}
\date{April 16, 2017}

\author{E.~V.~Gorbar}
\affiliation{Department of Physics, Taras Shevchenko National Kiev University, Kiev, 03680, Ukraine}
\affiliation{Bogolyubov Institute for Theoretical Physics, Kiev, 03680, Ukraine}

\author{V.~A.~Miransky}
\affiliation{Department of Applied Mathematics, Western University, London, Ontario, Canada N6A 5B7}

\author{I.~A.~Shovkovy}
\affiliation{College of Integrative Sciences and Arts, Arizona State University, Mesa, Arizona 85212, USA}
\affiliation{Department of Physics, Arizona State University, Tempe, Arizona 85287, USA}

\author{P.~O.~Sukhachov}
\affiliation{Department of Applied Mathematics, Western University, London, Ontario, Canada N6A 5B7}

\begin{abstract}
In this review we discuss a wide range of topological properties of electron quasiparticles in Dirac
and Weyl semimetals. Their nontrivial topology
is quantified by a monopole-like
Berry curvature in the vicinity of Weyl nodes, as well as by the energy and momentum space separations
between the nodes.
The momentum separation,
which is also known as the chiral shift, is one of the key elements of this review. We show that it
can be dynamically
generated in Dirac materials in a background magnetic field.
We also pay a special attention to various
forms of interplay between the background electromagnetic fields and the topological characteristics of Dirac
and Weyl semimetals. In particular, we discuss their signature features in
the transport of the electric and chiral charges, heat, as well as the quantum oscillations associated with the Fermi arc states. The origin
of the dissipative transport of the Fermi arc states is critically examined. Finally, a consistent chiral kinetic theory
for the description of Weyl semimetals is reviewed and its applications are demonstrated.
\end{abstract}

\maketitle
\tableofcontents

\newpage

\section{Introduction}
\label{sec:Introduction}

Although the chiral anomaly for relativistic fermions was discovered a long time ago~\cite{Adler,Bell-Jackiw},
it enjoyed a rather unexpected renaissance in recent years in diverse physical systems such
as heavy-ion collisions and the primordial plasma in the early Universe,
as well as Dirac and Weyl semimetals in condensed matter. This renaissance is due to the recent
understanding that a chiral asymmetry in relativistic matter in the presence of background electromagnetic
fields can lead to new anomalous transport phenomena.

Since the chirality plays a central role in this review, let us briefly remind what it is.
While a nonrelativistic fermion can be at rest and its spin can point in any direction,
a massless relativistic fermion always moves with the speed of light. Its
spin is strictly parallel to the particle's momentum and points either along or opposite to the direction of motion for the right- or left-handed fermions, respectively. Then, it is clear that the chiral asymmetry can be achieved in two ways. The simplest realization is connected with
the excess of fermions of one chirality.
Since the chiral charge density is given by $\bar{\psi}\gamma^0\gamma^5\psi$,
such an excess can be described by the term $b_0\bar{\psi}\gamma^0\gamma^5\psi$ in the fermion action. Here $\gamma^{\mu}$ are the standard gamma matrices, $\gamma^5=i\gamma^{0}\gamma^{1}\gamma^{2}\gamma^{3}$, and $\bar{\psi}=\psi^{\dag} \gamma^0$ is the Dirac conjugate spinor.
From a physics viewpoint, $b_0$ has a transparent meaning, i.e., it describes the splitting
between a pair of opposite-chirality Weyl nodes in energy and breaks the parity inversion (PI) symmetry.

Another form of chiral asymmetry can be achieved when there are equal number densities
of the right- and left-handed fermions, but they move in opposite directions. In this case, while the electric
current (which is the sum of the currents of the right- and left-handed fermions) vanishes, the axial
current (which is the difference of the corresponding currents) is nonzero. Note that the corresponding
axial current in a relativistic model could be viewed also as a spin density.
Mathematically, the axial current density is defined as $\bar{\psi}\bm{\gamma}\gamma^5\psi$. The
corresponding current appears in relativistic matter at a nonzero electric chemical potential
$\mu$ in a magnetic field $\mathbf{B}$ and is given by $\mathbf{J}_5 =-e^2\mu\mathbf{B}/(2\pi^2)$. This
phenomenon is known in the literature as the chiral separation effect (CSE)~\cite{Vilenkin:1980fu,Metlitski:2005pr,Newman:2005as}.
Note that this axial current stems exclusively from to the spin polarized lowest Landau level (LLL) and
is related to the chiral anomaly. Therefore, physically, the CSE could be considered as a form
of relativistic paramagnetism. Interestingly, if we add the term $\mathbf{b}\bar{\psi}\bm{\gamma}\gamma^5\psi$
to the Dirac Hamiltonian of massless fermions, the resulting model will contain two Weyl nodes
of opposite chirality separated by $\mathbf{b}$ in momentum space. For obvious reasons, we called
$\mathbf{b}$ the chiral shift in our paper~\cite{Gorbar:2009bm}. One can easily check that $\mathbf{b}$
breaks the time reversal (TR) symmetry. The chiral shift is one of the key players in this review.

From the viewpoint of high energy physics, parameters $b_0$ and $\mathbf{b}$ defining the chiral
asymmetry in relativistic matter are the time-like and space-like components of
the axial four-vector potential $b^{\nu}=(b_0,\mathbf{b})$ coupled to the axial current
$\bar{\psi}\gamma^{\nu}\gamma^5\psi$, where $\nu=\overline{0,3}$. In the presence of external
electromagnetic fields, the chiral asymmetry leads to new transport phenomena. In addition to the
CSE discussed above, the chiral magnetic effect (CME)~\cite{Kharzeev:2007tn,Kharzeev:2007jp,Fukushima:2008xe}
connected with the electric current $\mathbf{J}=-e^2\mu_5\mathbf{B}/(2\pi^2)$ in the background
magnetic field $\mathbf{B}$ is perhaps the most well-known.
Here $\mu_5$ is the chiral chemical potential that determines the imbalance between the fillings of opposite chirality Weyl nodes.
Among other anomalous
transport phenomena are the chiral electric separation effect (CESE)~\cite{Huang:2013}, the
chiral vortical effect (CVE)~\cite{Chen:2014cla}, etc.

Let us now proceed to the condensed matter systems where unusual chiral asymmetry
phenomena can be realized and, very importantly, can be observed by using rather simple
experimental techniques. These systems are Dirac and Weyl semimetals whose quasiparticle
excitations are described by the Dirac and Weyl equations, respectively. The study of these materials
has a long history. Already in 1937, Herring considered~\cite{Herring}
the problem of two bands touching
at isolated points in the Brillouin zone. He showed
that the Hamiltonian describing the electron states in vicinity of such points is naturally represented
by a $2\times2$ matrix, which coincides with the Weyl Hamiltonian in the case of a linear band crossing.
Later, this question was also studied by
Abrikosov and Beneslavskii \cite{Abrikosov-1:1971}, who showed that the symmetries in
some crystals could indeed allow for a linear spectrum. In addition, they considered the role of the Coulomb interaction that leads to a slowly varying logarithmic factor in the dynamical variables, including the vertex, polarization, and self-energy functions.

Theoretically, Dirac semimetals were first predicted to be realized in the A$_3$Bi (A=Na,K,Rb) and Cd$_3$As$_2$
compounds~\cite{WangWeng,Fang} in 2012. By using the angle-resolved photoemission spectroscopy (ARPES), the Dirac
semimetal band structure was indeed observed~\cite{Borisenko,Neupane,Liu} in $\mathrm{Cd_3As_2}$ and
$\mathrm{Na_3Bi}$. Further, the existence of Weyl semimetals was theoretically predicted in pyrochlore iridates
in Ref.~\cite{Savrasov}. In 2015 they were experimentally observed in the TaAs, TaP, NbAs,
and NbP compounds~\cite{Tong,Bian,Qian,Long,Xu-Hasan:TaP,Xu-Hasan:NbAs,Xu-Feng:NbP,Shekhar-Nayak:2015,Wang-Zheng:2015,Zhang-Xu:2015} (for a recent review regarding the discovery of the Weyl semimetals, see Ref.~\cite{Hasan-Huang:2017-Rev}).

Later, the materials with the topological charges of the Weyl
nodes greater than one, i.e., multi-Weyl semimetals, were also suggested to be realized in
$\mbox{HgCr}_2\mbox{Se}_4$~\cite{Xu-Fang:2011,Fang-Bernevig:2012} and $\mbox{SrSi}_2$~\cite{Huang}.
In accordance with the crystallographic point symmetries ~\cite{Fang-Bernevig:2012}, only the Weyl nodes with
topological charges $n_{\text{\tiny W}}$ less than or equal to $3$ are permitted. The corresponding double-Weyl ($n_{\text{\tiny W}}=2$)
and triple-Weyl ($n_{\text{\tiny W}}=3$) semimetals have the quadratic and cubic energy dispersion relations, respectively.
For recent reviews of Weyl semimetals, see
Refs.~\cite{Yan-Felser:2017-Rev,Hasan-Huang:2017-Rev,Armitage-Vishwanath:2017-Rev}.

Among the characteristic features of Dirac and Weyl semimetals are their unusual transport properties in
external electromagnetic fields, which are profoundly affected by the chiral anomaly. As was first shown by Nielsen
and Ninomiya~\cite{Nielsen}, the longitudinal (with respect to the direction of a magnetic field) magnetoresistivity
in Weyl semimetals decreases with the growth of the magnetic field. 
Therefore, this phenomenon
is called a negative magnetoresistivity in the literature. It was first experimentally observed in
$\mathrm{Bi_{1-x}Sb_x}$ alloy with $\mathrm{x} \approx 0.03$~\cite{Kim:2013dia},
and later in other Dirac and Weyl materials~\cite{Xiong-Ong:2015,Li-Wang:2015,Feng-Pang:2015,Liang-Ong:2015,Li-He:2015,Li-Valla:2016,Ong,Long,Shekhar-Nayak:2015,Wang-Zheng:2015,Zhang-Xu:2015}.

In addition to the anomalous bulk properties, Weyl semimetals possess also unusual surface states.
They are known as the Fermi arc states and provide a distinctive hallmark of these
materials.
Experimentally, the Fermi arcs are directly observed
with the help of the ARPES technique (see, e.g.,
Ref.~\cite{Hasan-Huang:2017-Rev} and references therein).
Unlike the long-established Tamm--Shockley~\cite{Tamm:1932,Shockley:1939} states that form
closed loops in the momentum space, the Fermi arcs are open segments of the Fermi surface connecting the projections
of the bulk cones onto the surface~\cite{Savrasov,Aji,Haldane}.
The existence of the surface Fermi arcs is directly related to the nontrivial
topological properties of Weyl nodes. Indeed, a Weyl node is a monopole of the Berry
curvature~\cite{Berry:1984} whose topological charge is connected with the chirality of the
node. Further, according to the Nielsen--Ninomiya theorem~\cite{Nielsen-Ninomiya-1,Nielsen-Ninomiya-2},
Weyl nodes in crystals always occur only in pairs of opposite chirality. Then, since vacuum has trivial
topological properties, the surface of a Weyl semimetal necessarily contains the topologically
protected surface Fermi arc states. As was emphasized in Ref.~\cite{Haldane}
these states are crucial for equilibrating the quasiparticles of opposite
chirality and maintaining the same value of the
chemical potential for the whole system.

This paper is organized as follows. In Sec.~\ref{sec:generation},
we discuss the mathematical origin of the chiral shift and its dynamical generation in Dirac semimetals subjected to
a background magnetic field. The bulk magnetotransport in Dirac and Weyl semimetals is discussed in
Sec.~\ref{sec:magnetotransport}. In Sec.~\ref{sec:Fermi-arcs}, the Fermi arc states in Weyl semimetals are
introduced and their physical implications for the quantum oscillations and the surface transport are discussed.
In Sec.~\ref{sec:strain}, we describe the bulk transport phenomena in strain-induced pseudoelectromagnetic
fields as well as formulate the quasiclassical kinetic framework in Weyl semimetals. Sec.~\ref{sec:BZ} is devoted to
the electric, chiral, and thermoelectric transport in a two-band model of Weyl semimetals.
The results are summarized in Sec.~\ref{sec:Summary}. Throughout the paper, we use the units with $\hbar = c = 1$.

\section{The dynamical generation of the chiral shift}
\label{sec:generation}

In this section, we discuss one of the defining parameters of Weyl semimetals, i.e., the chiral shift.
From a physics viewpoint, it determines the momentum space separation of the Weyl nodes. As we discuss
in Subsec.~\ref{sec:generation-Haldane}, such a parameter could be also considered as a 3D analog
of the Haldane gap in (2+1)-dimensional theories. In Subsec.~\ref{sec:generation-engineering}, we show
that the chiral shift can be dynamically generated in Dirac semimetals in a background magnetic field.

\subsection{Haldane mass in graphene as a prototype of the chiral shift}
\label{sec:generation-Haldane}

The experimental discovery of graphene~\cite{Geim2004Science}, whose electron quasiparticles are
described by the (2+1)-dimensional Dirac equation~\cite{Semenoff1984PRL},
strongly influenced condensed matter physics (see, e.g., Refs.~\cite{Neto-Geim:rev,Sarma:rev,Katsnelson:rev}).
Interestingly, the breakdown of the spin-lattice $U(4)$ symmetry of the graphene low-energy Hamiltonian was experimentally observed in a magnetic field (for reviews, see Refs.~\cite{Yango:2007,Goerbig:2011}).
One of the scenarios describing this symmetry breaking is the magnetic catalysis (see, e.g., Ref.~\cite{Shovkovy:rev} and references therein).
It was revealed that, in addition
to the standard PI and TR invariant Dirac mass, the TR breaking Haldane
mass~\cite{Haldane1988PRL} can be dynamically generated in graphene
in a strong magnetic field~\cite{Gorbar2008PRB,Gorbar2008LTP}. Historically, the Haldane mass was first discussed
in the context of $(2 + 1)$-dimensional relativistic gauge field theories~\cite{Niemi1983PRL,Redlich1984PRL},
where it induces the Chern--Simons term in the effective action.

By making use of the usual four-component Dirac spinor
$\Psi_s = \left( \psi_{KAs},\psi_{KBs},\psi_{K^\prime Bs},\psi_{K^\prime As}\right)$
that combines the Bloch states with given spin ($s=\uparrow,\downarrow$), sublattice ($A,B$), and valley
($K,K^\prime$), the explicit structure of the Haldane mass term reads $\Delta\bar{\Psi}\tilde{\gamma}^3\tilde{\gamma}^5\Psi$.
Here $\overline{\Psi}_{s}=\Psi^\dagger_{s}\tilde{\gamma}^0$ is the Dirac conjugated spinor,
$\tilde{\gamma}^{\nu}=\tilde{\tau}_3\otimes (\tau_3,i\tau_2,-i\tau_1)$ (with $\nu=0,1,2$) are the
$4\times4$ gamma matrices belonging to a reducible representation of the Dirac algebra in $2+1$
dimensions, and the Pauli matrices $\tilde{\tau}$ and $\tau$ act
in the valley and sublattice spaces, respectively.

As we stated in the Introduction, Dirac and Weyl semimetals are materials whose quasiparticles are also described by the relativistic-like (3+1)-dimensional equations. Therefore, it is natural to ask what is an analog of the Haldane mass in
3D case and whether it could be dynamically generated. Historically, these questions
were the turning point for the three of the current authors in the subsequent studies of the chiral
asymmetry in 3D relativistic matter~\cite{Gorbar:2009bm,GMSh2,Gorbar:2011ya}.

First of all, it is immediately clear that $\bar{\psi}\gamma^3\gamma^5\psi$ has a completely
different physical meaning in (3+1)-dimensional theories. It is no longer a mass term like in
graphene, but an axial-vector current density that breaks the TR symmetry.
By using the Nambu--Jona--Lasinio model with a local four-fermion interaction, we showed~\cite{Gorbar:2009bm} that the term
$b_z\bar{\psi}\gamma^3\gamma^5\psi$ in the effective fermion action could be dynamically generated in
dense relativistic matter when a magnetic field is present.
Further, by using the gauge invariant point-splitting regularization, we
checked~\cite{GMSh2} that although the chiral shift $b_z$ contributes to the axial current, it does not affect the chiral anomaly relation at all.

Concerning its interpretation, a word of caution should be given about $b_z$. While the Haldane
mass has a direct physical meaning in 2D, a constant axial-vector potential in 3D may not be observable {\it a priori}.
Indeed, naively it could be easily removed by the gauge transformation
$\psi\to e^{iz\gamma^5b_z}\psi$, $\bar{\psi} \to \bar{\psi}e^{iz\gamma^5b_z}$.
The point, however, is that this transformation is anomalous. This follows
from the two facts: (i) $b_z$ acts as a correction to the electric chemical potential on the LLL,
and (ii) the LLL dynamics is $(1+1)$-dimensional~\cite{MC2-1,MC2-2}. Since
the chiral transformation is anomalous in the $1 + 1$ dimensions (for a recent
discussion of this transformation, see Ref.~\cite{Kojo}), one cannot simply get rid of the constant
axial-vector potential in the presence of background electromagnetic fields.

Interestingly, we found~\cite{Gorbar:2011ya} that the dynamically induced $b_z$
is not inhibited by a nonzero temperature. This implies that the chiral shift can be also
generated in the regime relevant for heavy-ion collisions as well as in the cores of neutron stars,
where strongly magnetized quark matter could be present.
Clearly, the same should apply to the Dirac semimetals whose quasiparticles are described by the (3+1)-dimensional Dirac equation. This idea is further elaborated in the next subsection.

\subsection{Engineering Weyl nodes in Dirac semimetals by a magnetic field}
\label{sec:generation-engineering}

In this subsection, we consider an interesting property of Dirac and Weyl semimetals:
a dynamical rearrangement of their Fermi surfaces in a magnetic field~\cite{Gorbar:2013qsa}.
We begin with writing down the general form of the low-energy Hamiltonian for a Weyl semimetal
\begin{equation}
H^{\rm (W)}=H^{\rm (W)}_0+H_{\rm int},
\label{Hamiltonian-model-Weyl}
\end{equation}
where
\begin{equation}
H^{\rm (W)}_0=\int d^3r \left[\,v_F \psi^{\dagger} (\mathbf{r})\left(
\begin{array}{cc} \bm{\sigma}\cdot(-i\bm{\nabla} +e\mathbf{A} - \mathbf{b}_0)  & 0\\ 0 &
-\bm{\sigma}\cdot(-i\bm{\nabla} +e\mathbf{A} + \mathbf{b}_0) \end{array}
\right)\psi(\mathbf{r})-\mu_{0}\, \psi^{\dagger} (\mathbf{r})\psi(\mathbf{r})
\right]
\label{free-Hamiltonian}
\end{equation}
is the Hamiltonian of the free theory, which describes two Weyl nodes of opposite chirality
(as required by the Nielsen--Ninomiya theorem~\cite{Nielsen-Ninomiya-1,Nielsen-Ninomiya-2}) separated by vector $2\mathbf{b}_0$
in momentum space. Following Refs.~\cite{Gorbar:2009bm,Gorbar:2011ya},
$\mathbf{b}_0$ can be interpreted as the bare chiral shift parameter. There are two reasons for choosing such a
terminology. Firstly, as Eq.~(\ref{free-Hamiltonian}) implies, vector $\mathbf{b}_0$ shifts
the positions of Weyl nodes from the origin in the momentum space and, secondly, the
shift has opposite signs for the fermions of different chirality. The other notations are:
$\mu_{0}$ is the bare electric chemical potential, $\bm{\sigma}=(\sigma_x,\sigma_y,\sigma_z)$
are the Pauli matrices associated with the conduction-valence band degrees of freedom in
a generic low-energy model~\cite{ZyuzinBurkov}, which are commonly called the pseudospin
matrices, and $H_{\rm int}$ is the interaction part of the Hamiltonian. In addition, due to
the presence of an external magnetic field, we replaced $\bm{\nabla} \to \bm{\nabla}+ie\mathbf{A}$,
where $\mathbf{A}$ is the vector potential and $e$ is the absolute value of the electron charge.
The low-energy Hamiltonian of Dirac semimetals is trivially obtained from Eq.~(\ref{Hamiltonian-model-Weyl})
by setting $\mathbf{b}_0=\mathbf{0}$.

By assuming a simple model with a contact four-fermion interaction, i.e.,
$U(\mathbf{r}) = e^2/(\kappa |\mathbf{r}|) \rightarrow g\, \delta^3(\mathbf{r})$,
where $\kappa$ is a dielectric constant and $g$ is a dimensionful coupling constant,
and using the mean-field approximation, one can derive the following set of the Schwinger--Dyson (gap)
equations:
\begin{align}
\label{gap-mu-text}
\mu &= \mu_0 +\frac{3}{4e}\, g \,\langle J^{0}\rangle , \quad &&\langle J^{0}\rangle \equiv  e\,\mbox{tr}\left[\gamma^0G(u,u)\right],\\
\label{gap-tilde-mu-text}
\bm{\tilde{\mu}} &= \frac{1}{4}\, g \,\langle \bm{\Sigma} \rangle , \quad &&\langle \bm{\Sigma} \rangle \equiv -\mbox{tr}\left[ \gamma^0\bm{\gamma}\,\gamma^5G(u,u)\right],  \\
\label{gap-m-text}
m &= - \frac{1}{4}\, g \,  \langle \bar{\psi}\psi\rangle , \quad &&\langle \bar{\psi}\psi\rangle \equiv -\mbox{tr}\left[G(u,u)\right]\\
\label{gap-Delta-text}
\mathbf{b} &= \mathbf{b}_0 - \frac{1}{4e}\, g \, \langle\mathbf{J}_5\rangle, \quad &&\langle \mathbf{J}_5\rangle \equiv  e\,\mbox{tr}\left[\bm{\gamma}\,\gamma^5G(u,u)\right],
\end{align}
where, by definition, $G(u,u^{\prime})$ denotes the full fermion propagator with the following ansatz for the inverse one:
\begin{equation}
iG^{-1}(u, u^{\prime})= \Big[(i\partial_t+\mu )\gamma^0 - \left(\bm{\pi}\cdot\bm{\gamma}\right)
+\gamma^0(\bm{\tilde{\mu}}\cdot\bm{\gamma})\gamma^5
+(\mathbf{b} \cdot \bm{\gamma})\gamma^5
-m\Big]\delta^{4}(u-u^{\prime}).
\label{ginverse}
\end{equation}
Here $u=(t,\mathbf{r})$ and $\bm{\pi} \equiv -i \bm{\nabla} + e\mathbf{A}$ is the canonical momentum.
[The eigenvalues of $\gamma^5$,
i.e., $\lambda=\pm$, correspond to the node (chirality) degree of freedom.] Further, $m$ plays the role
of the dynamical Dirac mass.
As follows from the Dirac structure of the $\tilde{\bm{\mu}}$ term, it can be re-interpreted as
the anomalous magnetic moment $\mu_{\rm an}$, i.e., $\tilde{\bm{\mu}}\equiv \mu_{\rm an}\mathbf{B}$,
associated with the pseudospin.
It should be also emphasized that the renormalized $\mu$ and $\mathbf{b}$ in the full propagator may
differ from their tree-level counterparts $\mu_0$ and $\mathbf{b}_0$, respectively.
Further, $\langle J^{0} \rangle$ is equivalent to the electric charge density $\rho$ and $\langle \bm{\Sigma} \rangle$ describes
the pseudospin density of Weyl nodes. On the other hand, the chiral condensate
$\langle \bar{\psi}\psi\rangle$ and the axial current density $\langle \mathbf{J}_5\rangle$ describe internode coherent
effects related to a charge density wave and a valley polarized pseudospin condensate, respectively.

Because of the CME~\cite{Vilenkin:1980fu,Metlitski:2005pr,Newman:2005as}, the axial current density $\langle \mathbf{J}_5\rangle$ is
generated in the free theory when a fermion charge density and a magnetic field are present. Therefore, according to
Eq.~(\ref{gap-Delta-text}), the chiral shift $\mathbf{b}$ is induced already in the lowest order of the perturbation theory
even if $\mathbf{b}_0=\mathbf{0}$. As a result, a Dirac semimetal with a nonzero charge density is transformed into a Weyl
one, as soon as an external magnetic field is applied to the system.

In the leading order of the perturbation theory, the electric charge density $\langle J^{0}\rangle_0\propto -e^2B$ and the axial
current density $\langle \mathbf{J}_5\rangle_0 = -e^2\mathbf{B}\mu_0/(2\pi^2)$ are nonzero.
As to the chiral and the anomalous magnetic moment condensates, they vanish,
i.e., $\langle \bar{\psi}\psi\rangle_0 = 0$ and $\langle\bm{\Sigma} \rangle_0 = \mathbf{0}$.
Then, taking into account Eqs.~(\ref{gap-tilde-mu-text}) and (\ref{gap-m-text}),
we conclude that both the parameter $\tilde{\bm{\mu}}$ and the Dirac mass $m$
are zero in the perturbation theory. Therefore, only the electric chemical potential and the chiral shift are perturbatively renormalized. The latter is quantified by
\begin{equation}
\mathbf{b} = \mathbf{b}_0 + \frac{ge\mathbf{B}\mu_0}{8\pi^2}.
\label{engineering-b}
\end{equation}
The above equation implies that the dynamical contribution to the chiral shift either renormalizes the bare chiral shift (if it is
present in the Hamiltonian of the system) or generates it dynamically.

However, in accordance with the magnetic catalysis~\cite{Shovkovy:rev}
the ground state at vanishing $\mu_0$ should be characterized by a dynamically generated Dirac mass
$m_{\rm dyn}$ that breaks the chiral symmetry. In the weakly coupled regime $g\to 0$, in
particular, one finds
\begin{equation}
m_{\rm dyn}^2 =\frac{v_F^2|eB|}{\pi}\exp\left(-\frac{\Lambda^2}{g|eB|}\right),
\label{DiracMass}
\end{equation}
where $\Lambda$ is a ultraviolet
cutoff in the model at hand, which, for example, can be expressed through the lattice spacing $a$ as $\Lambda\simeq\pi/a$.
Such a vacuum state can withstand a finite stress for a sufficiently small chemical potential $\mu_0< m_{\rm dyn}$. When
$\mu_0$ exceeds a certain critical value $\mu_{\rm cr}$, the chiral symmetry is restored and the normal state with
a nonvanishing chiral shift parameter $\mathbf{b}$ in Eq.~(\ref{engineering-b}) is realized.
The free energies of the two types of states, i.e., the perturbative state with a nonzero chiral shift (and no Dirac mass) and
the nonperturbative state with a dynamically generated Dirac mass
(and no chiral shift) become
equal at about $\mu_0\simeq m_{\rm dyn}/\sqrt{2}$. This is somewhat analogous to the Clogston relation in
superconductivity~\cite{Clogston}. Note that the dynamical chiral
symmetry breaking was also studied in Weyl semimetals in
Refs.~\cite{Ran,Chao,Zhang-chiral,Sukhachov:2014bta,Sukhachov:2014-Coulomb}.

Therefore, the manifestation of the dynamical rearrangement
of the Fermi surfaces in Dirac semimetals subjected to an external magnetic field
is quite spectacular: a Dirac semimetal is transformed into a Weyl one.
Each Dirac node transforms into a pair of Weyl nodes separated by the dynamically
induced (axial) vector $2\mathbf{b}$, whose direction coincides with the direction of the
background magnetic field. The magnitude of the vector $\mathbf{b}$ is determined
by the strengths of the magnetic field and the interaction, as well as the quasiparticle charge density.

\section{Magnetoresistivity of Dirac and Weyl semimetals}
\label{sec:magnetotransport}

In this section, we discuss the implications of the relativistic-like dispersion relation, the chiral anomaly, and the chiral shift for the electric charge
transport in Dirac and Weyl semimetals in a magnetic field.

Remarkably, the study of the unusual magnetoresistance phenomena in condensed matter systems with a linear dispersion law began by Abrikosov~\cite{Abrikosov:1998} more
than 10 year before the theoretical proposal of Dirac and Weyl semimetals.
The idea was based on the assumption that nonstoichiometric silver chalcogenides
are basically gapless semiconductors with a linear energy spectrum and a very small carrier concentration.
Note that, at that time, the nontrivial properties of Dirac and Weyl semimetals were not appreciated neither experimentally nor theoretically. Consequently, the corresponding terminology was not established.
However, Abrikosov noted~\cite{Abrikosov:1998} that the electrons in his model are similar to charged neutrinos and could be
identified with Weyl fermions. Further, he found that the transverse magnetoresistivity increases linearly with magnetic field.
An important feature of the linear spectrum is that, unlike more
conventional cases, the quantum condition persists to rather
small magnetic fields and sufficiently high temperatures. In order to distinguish this phenomenon from other kinds of magnetoresistance
phenomena, Abrikosov suggested to call it the quantum magnetoresistance.

Further, in a remarkable paper~\cite{Nielsen},
Nielsen and Ninomiya realized that the chiral anomaly is responsible for pumping the electrons between the nodes of opposite chirality in Weyl semimetals
at a rate proportional to the scalar product of the applied electric and magnetic fields
$\mathbf{E}\cdot\mathbf{B}$. Such a pumping ultimately leads to the decrease of the longitudinal magnetoresistivity in Weyl semimetals with the growth of magnetic field.
This phenomenon is known in the literature as a negative magnetoresistivity.
For the first time, it was experimentally
observed in Dirac semimetal Bi$_{\rm 1-x}$Sb$_{\rm x}$ for ${\rm x} \approx 0.03$~\cite{Kim:2013dia} and was interpreted as an experimental signature of
a Weyl semimetal phase, where a single Dirac point splits into two Weyl
nodes with opposite chirality and the separation between the nodes in momentum space is
proportional to the applied field (see also Subsec.~\ref{sec:generation-engineering} and the references therein). The physical reason for the negative magnetoresistivity is quite
transparent \cite{Nielsen,Son} and connected with the spatial dimensional reduction $3 \to 1$ in the low-energy
dynamics dominated by the LLL. Such a dimensional reduction is a universal phenomenon,
taking place in the dynamics of charged fermions in a magnetic field~\cite{MC2-1,MC2-2}. Since the LLL states at each Weyl node are described by an
effective chiral 1D theory, they cannot backscatter on
impurities. Therefore, the corresponding contribution to the conductivity
is finite only due to the internode scattering processes. Since the
density of states on the LLL grows linearly with magnetic field, the LLL conductivity increases too.
If it dominates, then the total magnetoresistivity
decreases with magnetic field.

Recently, by making use of the Kubo formulas, three of us calculated the magnetoconductivity tensor and showed
that the negative magnetoresistivity can be realized in Dirac semimetals too~\cite{Gorbar:2013dha}.
In addition, it was shown that the special nature of the LLL plays a profound role also in the
anomalous Hall contribution to the transverse conductivity in Weyl semimetals.

In order to analyze the conductivity, we used the relativistic-like model with the Hamiltonian defined in Eq.~(\ref{free-Hamiltonian}). All calculations were performed in the quantum regime by using the Kubo linear-response theory, where the direct current conductivity tensor reads
\begin{equation}
\sigma_{nm}  = \lim_{\Omega\to 0}\frac{\mbox{Im}\,\Pi_{nm}(\Omega+i0; \mathbf{0})}{\Omega}
\label{sigma-ij}
\end{equation}
and is expressed through the Fourier transform of the current-current correlation function
\begin{equation}
\Pi_{nm}(\Omega;\mathbf{0}) =  e^2 v_F^2 T \sum_{l=-\infty}^{\infty}
\int \frac{d^3 \mathbf{p}}{(2\pi)^3} \mbox{tr} \Big[ \gamma_n \bar{G}(i\omega_l ;\mathbf{p})
\gamma_m \bar{G}(i\omega_l-\Omega; \mathbf{p})\Big].
\label{Pi_Omega_k}
\end{equation}
Note that this function is given in terms of the translation invariant part of the quasiparticle Green function in the magnetic field,
$n,m=(x,y,z)$, and $\sum_{l=-\infty}^{\infty}$ denotes the summation over the Matsubara frequencies.

Let us first analyze the case of the longitudinal (along the direction of the magnetic field) conductivity.
In such a case, it is instructive to extract the LLL contribution $\sigma_{zz}^{\rm (LLL)}$, which is given by the following exact result:
\begin{equation}
\sigma_{zz}^{\rm (LLL)} =\frac{e^2 v_F |eB|}{4 \pi^2 \Gamma_0},
\label{sigma33LLL}
\end{equation}
where $\Gamma_0$ is the quasiparticle decay width of the LLL.

\begin{figure}
\begin{center}
\includegraphics[width=.45\textwidth]{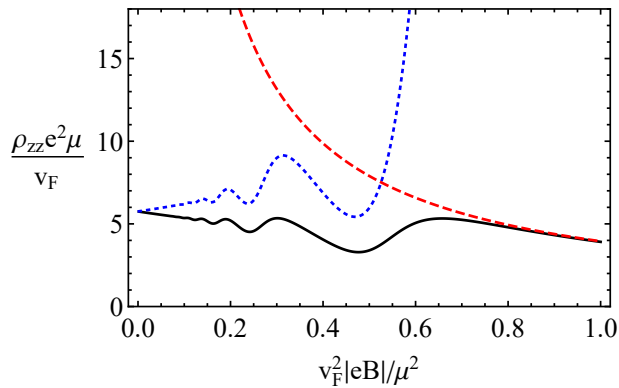}
\caption{Longitudinal resistivity $\rho_{zz}$ at zero temperature as a function of the magnetic field.
The solid black line shows the complete result, the red dashed line represents the contribution of the lowest Landau level,
and the blue dotted line describes the contribution due to the higher Landau levels.
The quasiparticle width is $\Gamma_0=\Gamma_n=0.1\mu$ in all Landau levels.}
\label{fig:resistivity-33}
\end{center}
\end{figure}

Taking into account that $\sigma_{xz}=\sigma_{zx}=\sigma_{yz}=\sigma_{zy}=0$, we easily find the longitudinal magnetoresistivity
$\rho_{zz}=1/\sigma_{zz}$. The corresponding numerical result including also the contributions of the higher Landau levels is plotted in
Fig.~\ref{fig:resistivity-33} as a function of $v^2_F|eB|/\mu^2$. The oscillations of
the magnetoresistivity connected with the Shubnikov-de Haas effect are clearly seen
in Fig.~\ref{fig:resistivity-33}. Due to the LLL contribution denoted by the dashed red line, the total longitudinal
magnetoresistivity decreases as the magnetic field grows. Therefore, both Dirac and Weyl semimetals exhibit the regime of the negative longitudinal magnetoresistivity at
sufficiently large magnetic fields. This result agrees with that obtained in Ref.~\cite{Nielsen} by using the chiral
anomaly in Weyl semimetals.
As a result, the experimental observation of the negative longitudinal magnetoresistivity cannot be used alone as an unambiguous signature of a Weyl semimetal.

In order to clarify this point, we note that the chiral shift $\mathbf{b}\parallel\mathbf{B}$ does not
influence the longitudinal magnetoconductivity and could affect the result
only indirectly through the quasiparticle width~\cite{Nielsen}. Moreover, as one can see in Fig.~\ref{fig:resistivity-33} (see Ref.~\cite{Gorbar:2013dha} for a detailed analysis),
the
negative longitudinal magnetoresistivity takes place even when the LLL quasiparticle width
$\Gamma_0$ is comparable to the width $\Gamma_n$ in the higher Landau levels. Therefore, we conclude that
this phenomenon is quite robust and takes place even as $b_z\to0$.

Next, let us analyze the transverse (with respect to the direction of the magnetic field) magnetotransport. While the diagonal components of the
transverse conductivity are associated exclusively with a nonzero density of charge carriers, the off-diagonal ones contain a topological
contribution (which is present even at vanishing electric chemical potential $\mu=0$ and temperature $T=0$) and comes exclusively from the completely occupied
states below the Fermi level. In the presence of an external magnetic field, such a term is related only to the LLL contribution.
It is a specific feature of Weyl semimetals and is directly related to the
anomalous Hall effect (AHE)~\cite{Ran,Burkov:2011ene,Grushin-AHE,Goswami,Burkov-AHE:2014}, which is produced by the dynamical Chern--Simons term in
Weyl semimetals~\cite{Burkov:2011ene,Grushin,ZyuzinBurkov,Franz,Basar:2014,Goswami} and related to the Bardeen--Zumino terms~\cite{Bardeen-1,Bardeen-2}
(see, also, Refs.~\cite{Landsteiner:2013sja,Landsteiner:2016}).
This topological (anomalous) contribution is independent of the magnetic field and equals
\begin{equation}
\sigma_{xy,\text{\tiny{AHE}}} = -\frac{e^2 b_z}{2\pi^2}.
\label{anomaly-contribution}
\end{equation}
Physically, it is manifested via the anomalous part of the electric current
$\mathbf{J}_{\text{\tiny{AHE}}}=e^2/(2\pi^2)[\mathbf{b}\times\mathbf{E}]$, which is similar to the Hall effect current where the role of a magnetic field is played by the chiral shift.
It worth noting that while one should treat the integrals over momentum with care in the linearized model,
the same topological result is straightforwardly reproduced in lattice models of Weyl semimetals.
As we will show in Subsec.~\ref{sec:BZ-I}, the topological origin of contribution (\ref{anomaly-contribution}) can be established by expressing it
in terms of a winding number in a lattice Hamiltonian model.

It should be noted that there is no interference between the topological contribution (\ref{anomaly-contribution}) and its matter counterparts, which are related to the finite density of charge carriers.
In addition, the former could be present even in Dirac semimetals when a magnetic
field is applied, because, as we discussed in Subsec.~\ref{sec:generation-engineering}, a nonzero chiral shift $\mathbf{b}$ can be generated
dynamically~\cite{Gorbar:2013dha}. Consequently, the anomalous contribution
(\ref{anomaly-contribution}) unambiguously distinguishes a Weyl semimetal from a Dirac one only in the absence
of a magnetic field.

\begin{figure}
\begin{center}
\includegraphics[width=.45\textwidth]{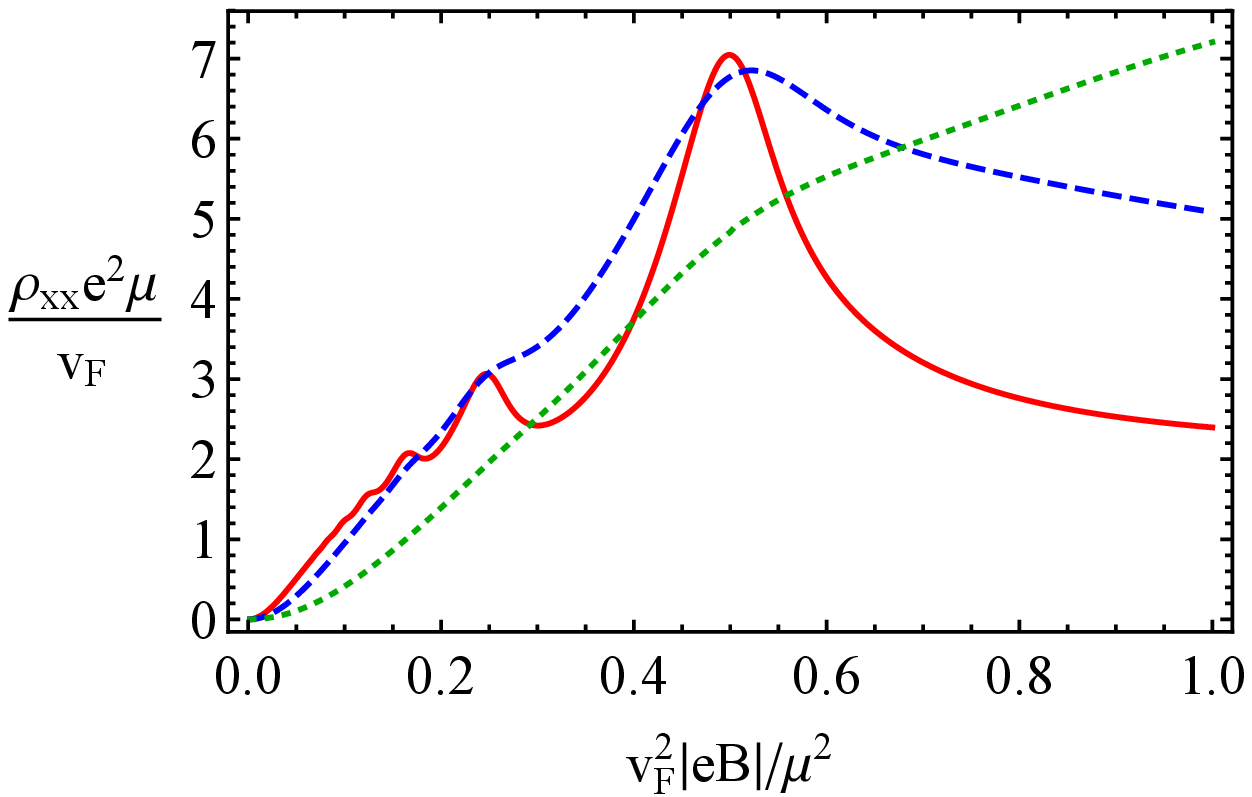}
\hfill
\includegraphics[width=.45\textwidth]{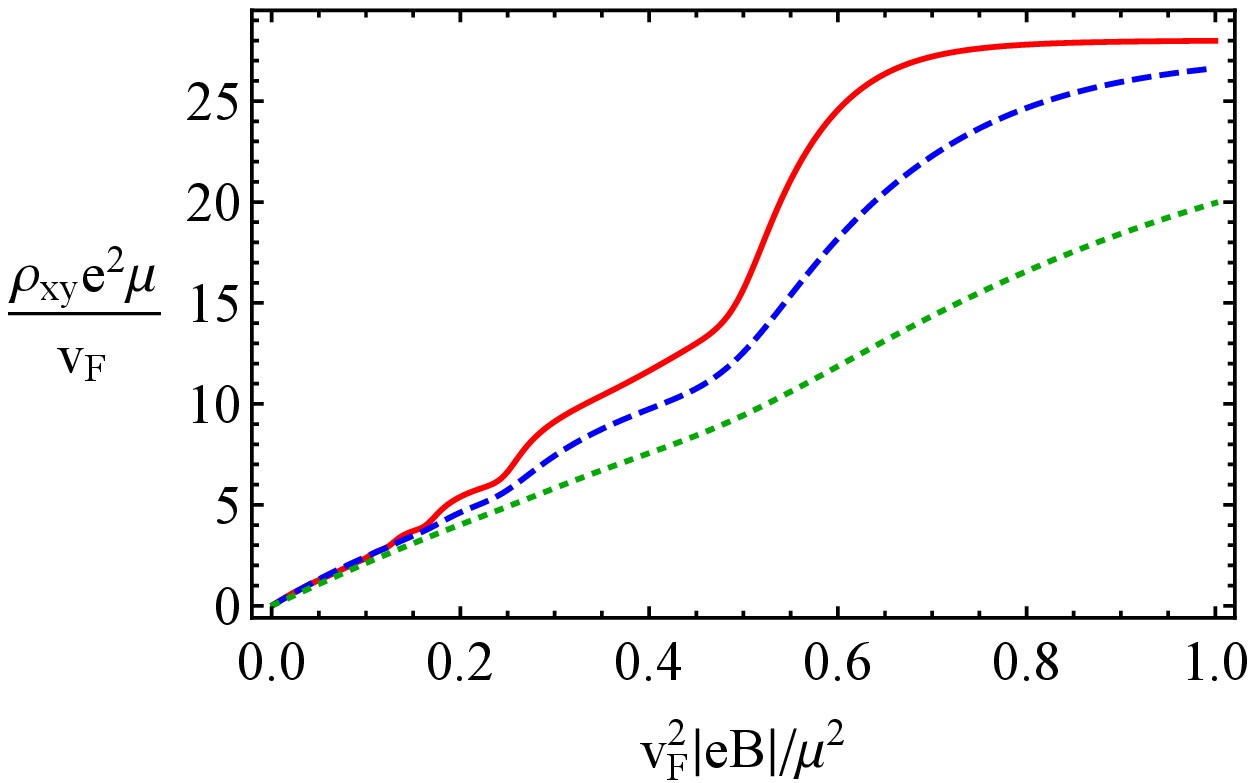}\\
\caption{Transverse components of the magnetoresistivity tensor $\rho_{xx}$ and $\rho_{xy}$ at zero temperature
as functions of the magnetic field for $b=0.3 \mu$.
The quasiparticle width is $\Gamma=0.05\,\mu$ (red solid line),
$\Gamma=0.1\,\mu$ (blue dashed line), and $\Gamma=0.2\,\mu$ (green dotted line).
}
\label{fig:resistivity-12}
\end{center}
\end{figure}

Further, we calculate numerically the transverse components of the magnetoresistivity tensor at zero temperature
\begin{equation}
\rho_{xx} = \rho_{yy} =\frac{\sigma_{xx}}{\sigma_{xx}^2+\sigma_{xy}^2}, \quad
\rho_{xy} =
\left|\frac{\sigma_{xy}}{\sigma_{xx}^2+\sigma_{xy}^2}\right|
\end{equation}
and plot them as functions of $v^2_F|eB|/\mu^2$ in Fig.~\ref{fig:resistivity-12}
for $|\mathbf{b}|\equiv b=0.3\, \mu$. As is clear from Figs.~\ref{fig:resistivity-33} and \ref{fig:resistivity-12}, all components of resistivity have the characteristic Shubnikov-de Haas oscillations
when the Landau levels are well resolved (i.e., the quasiparticle widths are sufficiently small). In passing, let us note that the magnetotransport in Weyl and Dirac semimetals was also studied in Refs.~\cite{Ominato:2013,Biswas:2013,Son-Spivak,Shen:2014,Syzranov:2014,Shen:2015,Spivak:2016}
(for recent reviews, see, e.g., Refs.~\cite{Lu-Shen-rev:2017,Wang-Lin-rev:2017}).

\section{Surface Fermi arc states and their transport properties}
\label{sec:Fermi-arcs}

In this section, we continue to study the anomalous transport properties of the Weyl and Dirac semimetals focusing
on 
their topologically nontrivial surface states, i.e., the Fermi arcs. We will start from introducing the
concept of the surface Fermi arcs in Subsec.~\ref{sec:Fermi-arcs-concept} and then discuss their physical implications
in Subsecs.~\ref{sec:Fermi-arcs-quantum-oscillations} and \ref{sec:Fermi-arcs-transport}.

\subsection{The Fermi arcs states in Weyl semimetals}
\label{sec:Fermi-arcs-concept}

In order to clarify the concept of the Fermi arcs, let us employ
one of the simplest
continuum models of a TR symmetry breaking Weyl semimetal (see, e.g., Ref.~\cite{Murakami}), whose Hamiltonian reads
\begin{equation}
\label{Hamiltonian-effective-M1}
H(\mathbf{k})=\gamma\left(k_z^2 - m\right)\sigma_z+v_F\left(k_x\sigma_x+k_y\sigma_y\right),
\end{equation}
where $m$ and $\gamma$ are positive constants. As is easy to check, the energy
spectrum of the bulk states is given by $\epsilon_{\mathbf{k}}=\pm\sqrt{\gamma^2(k_z^2-m)^2+v_F^2(k_x^2+k_y^2)}$
and the two Weyl nodes are located at $\mathbf{k}=(0,0,\pm \sqrt{m})$.

Further, in order to obtain the surface states, we need to define the appropriate boundary conditions.
We assume that, in coordinate space, the semimetal is in the upper
half-plane $y>0$ and the vacuum is in the lower half-plane $y<0$. In order to prevent the quasiparticles
from escaping into the vacuum region, we set $m\to -\tilde{m}$ and take the limit $\tilde{m}\to \infty$ for
$y<0$. Because of the semimetal surface at $y=0$, the translation invariance in the $y$ direction is
broken and, consequently, one should use the operator form $k_y= -i\partial_y$
in Hamiltonian (\ref{Hamiltonian-effective-M1}).

\begin{figure}[t]
\begin{center}
\includegraphics[width=0.45\textwidth]{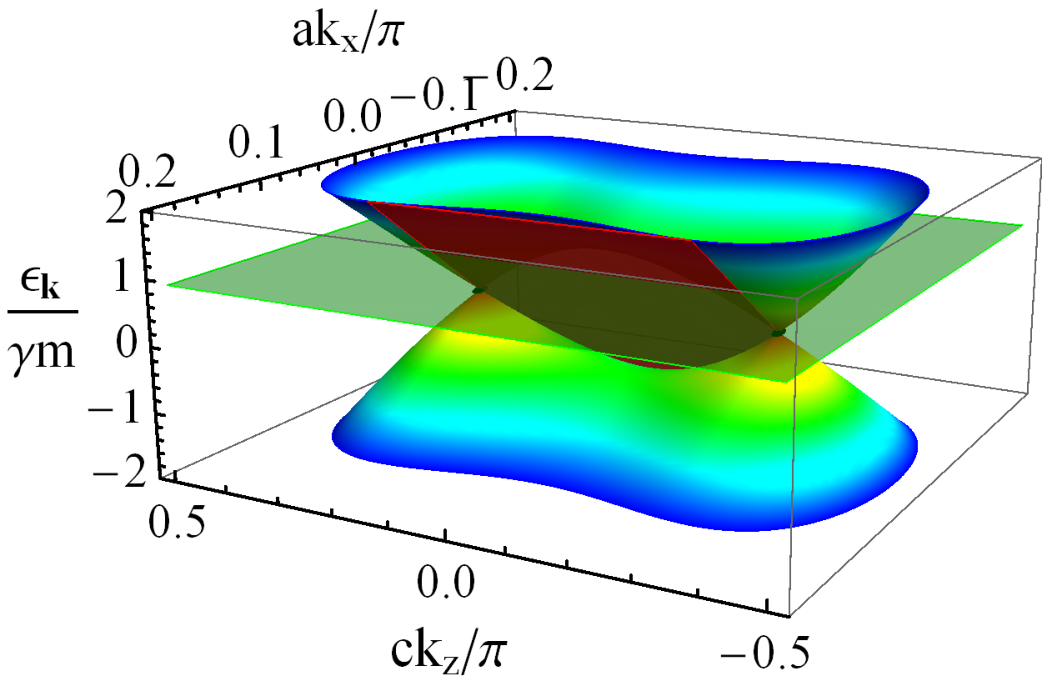}
\hfill
\includegraphics[width=0.45\textwidth]{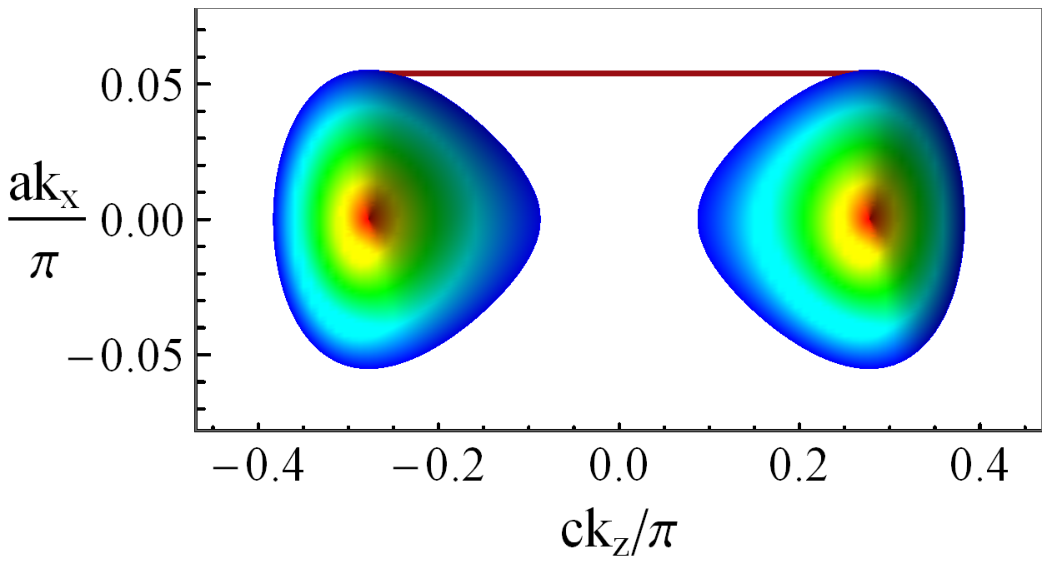}
\end{center}
\caption{The bulk and Fermi arc (red plane) states in the model of Weyl semimetal (\ref{Hamiltonian-effective-M1}) with $k_y=0$ (left panel). The green plane represents the
Fermi energy $\epsilon_F=0.9\,\gamma m$. The corresponding Fermi surface is plotted in the right panel. Here $a$ and $c$ are lattice spacings.}
\label{fig:Fermi-surf}
\end{figure}

By considering the eigenvalue problem $H\psi=\epsilon\psi$ and matching the wave functions at the boundary, it is not
difficult to find the following expressions for the surface states in the semimetal and vacuum:
\begin{eqnarray}
\psi_{y>0} &=& \sqrt{p(k_z)}\, e^{ik_xx+ik_zz-p(k_z)y}\left(
                                                              \begin{array}{c}
                                                                1 \\
                                                                1 \\
                                                              \end{array}
                                                            \right),
\label{psi-surface-states}
                                                            \\
\psi_{y<0} &=& \sqrt{p(k_z)}\, e^{ik_xx+ik_zz+\gamma \tilde{m}y}\left(
                                                              \begin{array}{c}
                                                                1 \\
                                                                1 \\
                                                              \end{array}
                                                            \right),
\label{psi-All-M1}
\end{eqnarray}
respectively, where $p(k_z)= \gamma (m-k_z^2)/v_F$ and $-\sqrt{m}<k_z<\sqrt{m}$. The latter condition follows from the normalizability of the wave functions.
The corresponding expression for the surface state energy reads
\begin{equation}
\epsilon_s= v_F k_x.
\label{energy-surface}
\end{equation}
Then, the effective Hamiltonian for the surface Fermi arc states is given by (see Ref.~\cite{Shen} for a similar derivation of the effective
Hamiltonian for the surface states in topological insulators)
\begin{equation}
H_{\rm surf}=-iv_F\partial_x,
\label{Hamiltonian-effective-surf-M1}
\end{equation}
and, as expected, describes one-dimensional chiral fermions. It is worth noting, however, that the effective Hamiltonian (\ref{Hamiltonian-effective-surf-M1}) is valid only for $-\sqrt{m}<k_z<\sqrt{m}$.
The energy spectrum of model (\ref{Hamiltonian-effective-M1}) as well as the Fermi arc states are
schematically shown in Fig.~\ref{fig:Fermi-surf}.

\subsection{Quantum oscillations in Weyl semimetals in a magnetic field}
\label{sec:Fermi-arcs-quantum-oscillations}

As we showed in the previous subsection, the Fermi arcs serve as a distinguishing fingerprint of the topologically nontrivial Weyl semimetals. While their
observation by using the ARPES technique is straightforward (see, e.g., Ref.~\cite{Hasan-Huang:2017-Rev} and references therein), it is important that
there are additional
transport means to investigate these unusual surface states~\cite{Potter:2014}. They are related to one of the oldest methods to investigate the Fermiology of
materials, i.e., the quantum oscillations of the density of states (DOS) in a magnetic field due to the closed orbits (see, e.g.,
Ref.~\cite{Shoenberg}). Then, according to the
Onsager relation, the period of the oscillations provides an information on the extremal cross-sections of the Fermi surface.
On the other hand, in a TR symmetry broken Weyl material, the Fermi arcs are disjoined and, naively, cannot contribute to the oscillations.
However, it was suggested in Ref.~\cite{Potter:2014} that
the closed magnetic orbits can indeed form in a magnetic field when the bulk states are taken into account.

\begin{figure}[!ht]
\begin{center}
\hspace{-0.32\textwidth}(a)\hspace{0.32\textwidth}(b)\hspace{0.32\textwidth}(c)\\[0pt]
\includegraphics[width=0.32\textwidth]{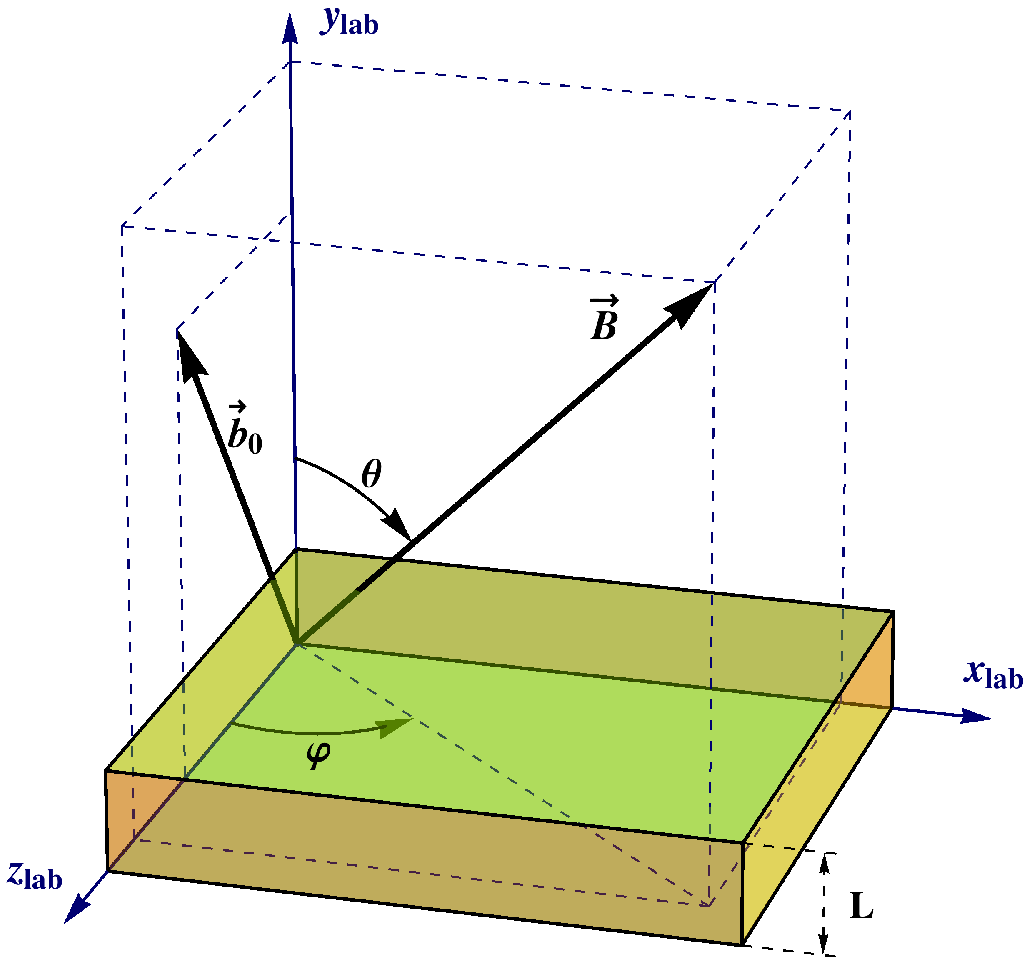}\hfill
\includegraphics[width=0.33\textwidth]{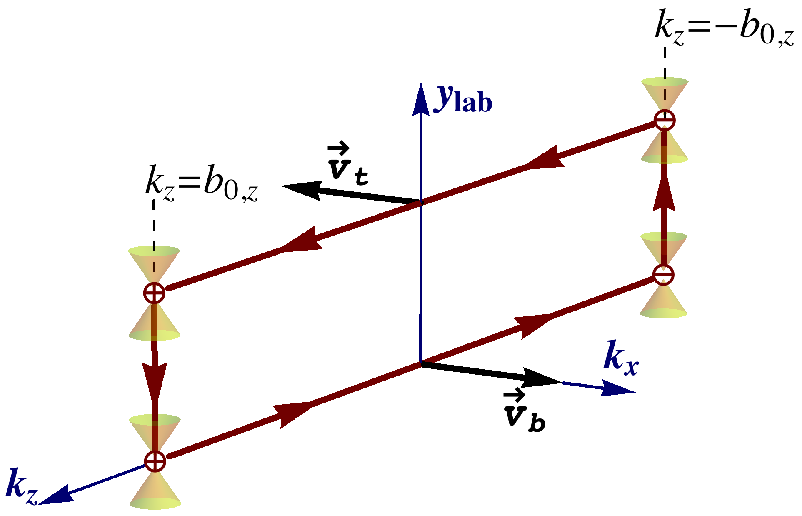}\hfill
\includegraphics[width=0.32\textwidth]{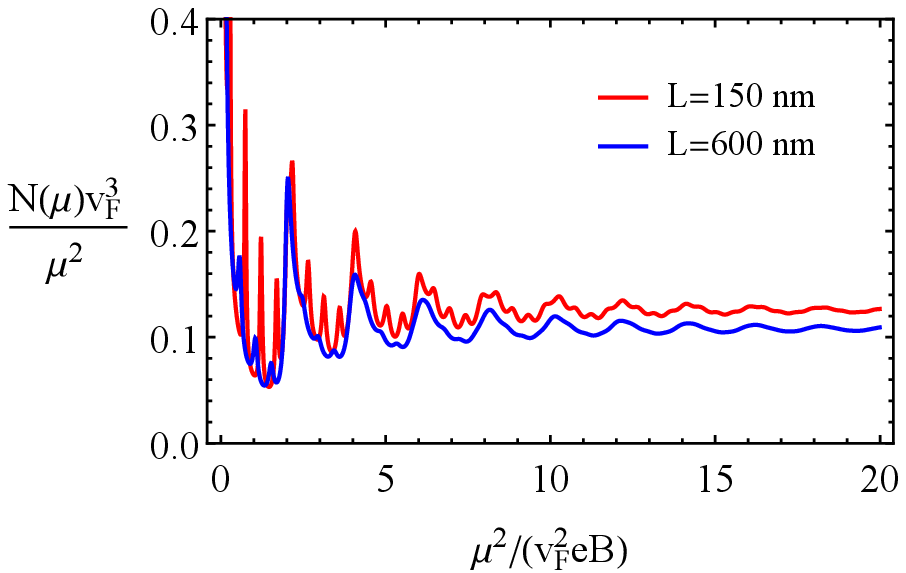}
\caption{The model setup (panel a) and a schematic representation of the closed
quasiparticle orbits in a magnetic field involving the surface Fermi arcs (panel b).
The oscillations of the total density of states including the bulk and surface contributions (panel c).}
\label{fig:oscillations}
\end{center}
\end{figure}

Following Ref.~\cite{Potter:2014}, let us briefly explain the physics behind this effect.
We start by assuming that the Weyl semimetal has the form of slab with the thickness $L$ and is characterized by the
Weyl nodes located at $\pm\mathbf{b}_{0}=\pm(0, b_{0,y}, b_{0,z})$, where $b_{0,y}$ and $b_{0,z}$ are the components of the
chiral shift perpendicular and parallel to the slab's surface, respectively.
As follows from Eq.~(\ref{energy-surface}) (see also Ref.~\cite{Gorbar:2014qta} for a detailed derivation of the
Fermi arcs in the slab geometry) the quasiparticle velocity of the Fermi arc modes is $\mathbf{v}_{\rm b}=(v_F, 0, 0)$
on the bottom surface ($y=0$) and $\mathbf{v}_{\rm t}=-\mathbf{v}_{\rm b}$ on the top surface ($y=L$).
The direction of the magnetic field $\mathbf{B}=B\, \hat{\mathbf{n}}$ is specified by the unit
vector $\hat{\mathbf{n}} = (\sin\theta\sin\varphi,\cos\theta,\sin\theta\cos\varphi)$, where the meaning of the angles $\theta$ and $\varphi$ is
obvious from Fig.~\ref{fig:oscillations}(a). The Fermi arc states are localized on the surfaces of the semimetal and are characterized by
wave vectors $k_x$ and $k_z$.

Further, let us describe the closed orbits schematically shown in Fig.~\ref{fig:oscillations}(b).
Since the velocities of the surface modes in the given framework are parallel to the $x$-axis,
only the $z$-component of the quasiclassical equation of motion in the magnetic field is
nontrivial, i.e.,
\begin{equation}
\partial_t k_{z} = -e\left[\mathbf{v}_{\rm b, t}\times\mathbf{B}\right]_{z}= \mp
e v_F B \cos{\theta},
\label{k_z-bottom-top}
\end{equation}
where the signs $\mp$ correspond to the top and bottom surfaces, respectively. Then, according to the above equation,
the quasiparticles slide along the bottom Fermi arc from the right-handed Weyl node at $k_z=b_{0,z}$  to the left-handed one at $k_z=-b_{0,z}$.
In the Weyl node, by using the gapless bulk states in the magnetic field as a conveyor belt, quasiparticles propagate to the top surface where
they move along the top Fermi arc, albeit in the opposite direction (which stems from the relation $\mathbf{v}_{\rm t}=-\mathbf{v}_{\rm b}$). Arriving at the
Weyl node, the quasiparticles move to the other surface, consequently, completing the orbit.

The semiclassical quantization condition for the closed orbits depicted in Fig.~\ref{fig:oscillations}(b) is
$E_n t = 2\pi(n+\alpha)$,
where $n$ is an integer and $\alpha$ denotes a phase shift that cannot be determined semiclassically.
Further, the time to complete the orbit $t$ includes the time of quasiparticles propagation through the bulk $t_{\rm bulk}=2 L/(v_F \cos{\theta})$ (where the presence of $\cos{\theta}$ in the denominator is related to the fact that the movement through the bulk occurs only along the magnetic field) and the time of the propagation along the arcs $t_{\rm arcs}$.
According to Eq.~(\ref{k_z-bottom-top}), the latter can be estimated as $t_{\rm arcs}=4b_{0,z}/(v_F e B \cos{\theta})$. Therefore, the quantization condition reads \cite{Potter:2014}
\begin{equation}
E_n = \frac{\pi v_F(n+\alpha)}{L/\cos{\theta}+2b_{0,z}/(e B \cos{\theta})}.
\label{semiclassical-condition-01}
\end{equation}
At fixed Fermi energy $\mu$, the maxima of the DOS oscillations are achieved at the following discrete values of the magnetic
field:
\begin{equation}
\frac{1}{B_n} = \frac{e}{2b_{0,z}} \left(\frac{\pi v_F \cos{\theta}}{\mu}(n+\alpha)-L\right).
\label{magnetic-inverse-g0-1}
\end{equation}
The period of these oscillations is
\begin{equation}
T_{1/B}=\frac{e\pi v_F \cos{\theta}}{2\mu b_{0,z}}.
\label{period-oscillations}
\end{equation}
The requirement that the right-hand side of Eq.~(\ref{magnetic-inverse-g0-1}) be positive defines the smallest possible value of $n$,
i.e., $n_{\rm min}=\left[ \mu L/(\pi v_F \cos{\theta})-\alpha+1\right]$, where $[\ldots]$ denotes the
integer part. It has a transparent physical meaning and defines the saturation value of the magnetic field
$B_{\rm sat} \equiv B_{n_{\rm min}}$. The oscillations of the DOS could be observed only for $B<B_{\rm sat}$.

As we discussed in Subsec.~\ref{sec:generation-engineering}, the interaction effects change the separation between the Weyl nodes
and, consequently, may also affect the quantum oscillations. In particular, the chiral shift vector is renormalized and
reads as $\mathbf{b}=\mathbf{b}_{0}+\delta{b}\, \hat{\mathbf{n}}$ (see Ref.~\cite{Gorbar:2014qta} for the detailed derivation). Here $\delta{b}$
is the magnitude of the correction to the chiral shift, which is similar to that in Eq.~(\ref{engineering-b}), and $\hat{\mathbf{n}}$ is the unit vector pointing in the direction of the magnetic field.
The effect of the interactions can be easily included by replacing the bare
arc length $2b_{0,z}$ with
$2\sqrt{\left(b_{0,z}\right)^2 + 2b_{0,z} \delta{b} \cos\varphi  \sin\theta +\left(\delta{b} \sin\theta \right)^2}$. As one can see, the most
important qualitative effect is the $\varphi$-dependence of the oscillation period. As we suggested in Ref.~\cite{Gorbar:2014qta}, this
dependence could be used to extract the dynamically generated correction to the chiral shift in Weyl semimetals.

Since Weyl semimetals have the gapless energy spectrum, the conventional oscillations in a magnetic field related to the extremal cross-section of
the bulk Fermi surface will be always present at any finite $\mu$~\cite{Ashby:2013}.
In order to study the relative contribution of the surface-bulk orbits to the total DOS $N(\mu)$, we present the latter in
Fig.~\ref{fig:oscillations}(c) for two values of the slab thickness $L$.
As one can see, the contribution of the surface-bulk orbits, which is given by thin peaks with a smaller period, strongly diminishes with $L$
leaving only the bulk background. Therefore, in order to reliably distinguish bulk and surface oscillations experimentally, one needs to use sufficiently thin samples and minimize the scattering effects, which tend to destroy the coherent electron motion.

In passing, let us discuss the experimental confirmation of the surface-bulk oscillations reported in Ref.~\cite{Moll:2016}.
The authors observed the two types of the magnetoresistance quantum oscillations: (i) the bulk type, which is almost independent on the
direction of the magnetic field, and (ii) the surface type with the period $\propto\cos{\theta}$. The amplitude of the latter strongly
depends on the slab's thickness $L$ (like shown in Fig.~\ref{fig:oscillations}(c)).
In order to exclude the contribution of the trivial surface states, where the electrons move in closed orbits confined to a particular
surface, the measurements of the samples of different geometries were also performed in Ref.~\cite{Moll:2016}. It was found that the
samples of a triangular size allow only for the bulk type of the magnetoresistance oscillations. This fact could be interpreted as the
destructive interference of oscillations with random phases.
The above findings provide a strong evidence in favor of the hybrid surface-bulk magnetic orbits.

\subsection{Origin of dissipative Fermi arc transport in Weyl semimetals}
\label{sec:Fermi-arcs-transport}

Taking into account the existence of the unusual surface-bulk orbits discussed in the previous subsection, it is a natural question how the
Fermi arcs contribute to the charge transport.
In view of the topological protection of these surface states in Weyl semimetals, it is not surmising
that their
effective Hamiltonian is that of a chiral 1D fermion
(see, e.g., Eq.~(\ref{Hamiltonian-effective-surf-M1})).
In this connection, it would
be useful to recall the quantum Hall effect~\cite{Klitzing}. There,
the absence of backscattering for chiral 1D fermions provides the physical reason for the nondissipative electric transport, resulting in
vanishing conductivity $\sigma_{xx}$. It was shown in Ref.~\cite{Kohmoto} that the fantastic
exactness of the QHE conductivity is connected with nontrivial topological properties of QHE materials, which are encoded in their nonzero
Chern numbers. Since vacuum has trivial topological characteristics, the current-carrying edge states are topologically protected that leads to
the celebrated bulk-boundary correspondence. All this strongly suggests that the electron surface transport due to the Fermi arcs is
nondissipative too.

The problem of the surface transport in Weyl semimetals was studied by us in Ref.~\cite{Gorbar:2016aov}.
We found that the
Fermi arc transport is, in fact, dissipative. The fundamental physical reason for the dissipation is the presence of gapless
bulk states in Weyl semimetals whose low-energy dynamics is not fully decoupled from the surface Fermi
arc states. In order to show this, we used a simple model of a Weyl semimetal in Eq.~(\ref{Hamiltonian-effective-M1}).
Recall that, in this model, there are two Weyl nodes located at $\mathbf{k}=(0,0,\pm \sqrt{m})$ that are connected by the Fermi arc with the energy
dispersion $\epsilon_s= v_F k_x$.
Further, by including a short-ranged quenched disorder, we showed~\cite{Gorbar:2016aov} that
in the full effective model of Weyl semimetals, which contains both the
surface Fermi arc states and the gapless bulk states, the Fermi arc transport is dissipative.
The formal reasons for this are: (i) the loss of the 1D kinematic
constraint in the scattering of chiral fermions and (ii) the nondecoupling of the low-energy dynamics
of the surface and bulk states. The latter stems from the fact that the quasiparticles from the surface Fermi arc states can scatter into the
bulk as well as into other Fermi arcs. This nondecoupling implies that there is no well-defined effective theory of the Fermi
arcs in Weyl semimetals in the presence of quenched disorder.
This is in contrast to the case of topological insulators, in which the bulk states are gapped. In addition, these findings agree with the
generic conclusions in Refs.~\cite{QPI:Mitchell, QPI:Derry}, claiming
that an effective theory of surface states in gapless systems can not be formulated.

Due to the loss of exact integrability
in the full model, we cannot present an exact solution supporting the nondissipativity of the Fermi arc transport.
However, it is possible to use physical arguments that strongly support such a claim.
Following Ref.~\cite{Gorbar:2016aov}, let us consider the scattering of the Fermi arc states using the Born approximation.
A convenient starting point for such an analysis is the Lippmann--Schwinger equation
for the surface states
\begin{equation}
\psi_s(\mathbf{r})=\psi^{(0)}_s(\mathbf{r})-i\int d^3\mathbf{r}^{\prime}
S(\mathbf{r}, \mathbf{r}^{\prime})U(\mathbf{r}^{\prime})
\psi_s(\mathbf{r}^{\prime} ),
\label{LP-equation}
\end{equation}
where $\psi^{(0)}_s(\mathbf{r})\equiv\psi_{y>0}$ is the incident surface wave, $S(\mathbf{r}, \mathbf{r}^{\prime})$ is the free propagator,
and $U(\mathbf{r}^{\prime})$ is the potential of impurities.
By making use of the perturbative approach one can use the following general definition of the propagator in a clean limit:
$S(\mathbf{r}, \mathbf{r}^{\prime})\equiv i\sum_E \psi_E(\mathbf{r})
\psi_E^{\dagger}(\mathbf{r}^{\prime})/(\omega-E+i0)$,
where the sum runs over a complete set of the energy eigenstates.
Further, we note that the Hilbert space contains both bulk and surface states. Therefore, the resulting Green function
naturally splits into the surface $S_{s}(\mathbf{r}, \mathbf{r}^{\prime})$ and bulk $S_{b}(\mathbf{r}, \mathbf{r}^{\prime})$ contributions,
leading to the following scattered wave function in the first order in the disorder potential:
\begin{eqnarray}
\label{psi-1}
\psi^{(1)}(\mathbf{r}) &=& \psi^{(1)}_s(\mathbf{r})+\psi^{(1)}_b(\mathbf{r}) , \\
\label{psi-1-s}
\psi^{(1)}_s(\mathbf{r}) &\simeq & -i\int d^3\mathbf{r}^{\prime} S_{s}(\mathbf{r}, \mathbf{r}^{\prime})
U(\mathbf{r}^{\prime})\psi^{(0)}_s(\mathbf{r}^{\prime} )
\simeq -i  u_{0} \sum_{j} S_{s}(\mathbf{r}, \mathbf{r}_{j})  \psi^{(0)}_s(\mathbf{r}_{j} ).
\end{eqnarray}
Here in $\psi^{(1)}_b(\mathbf{r})$ one needs to replace $S_{s}(\mathbf{r}, \mathbf{r}^{\prime})$ with $S_{b}(\mathbf{r}, \mathbf{r}^{\prime})$.
Obviously, the scattered wave functions consist of the surface and bulk contributions.
In addition, we used the local disorder potential $U(\mathbf{r}) =  \sum_j u_0 \delta(\mathbf{r}-\mathbf{r}_j)$ whose strength is determined by $u_0$.

\begin{figure}[t]
\begin{center}
\hspace{-0.32\textwidth}(a)\hspace{0.32\textwidth}(b)\hspace{0.32\textwidth}(c)\\[0pt]
\includegraphics[width=0.32\textwidth]{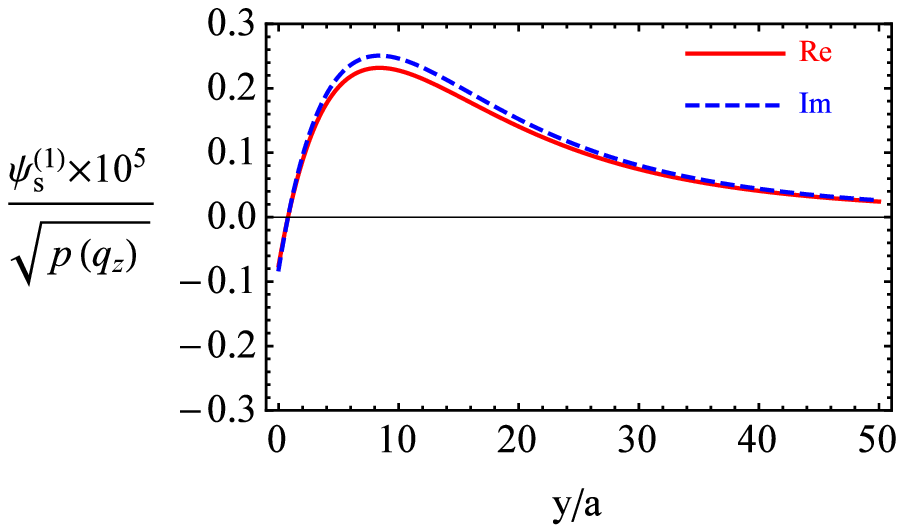}\hfill
\includegraphics[width=0.32\textwidth]{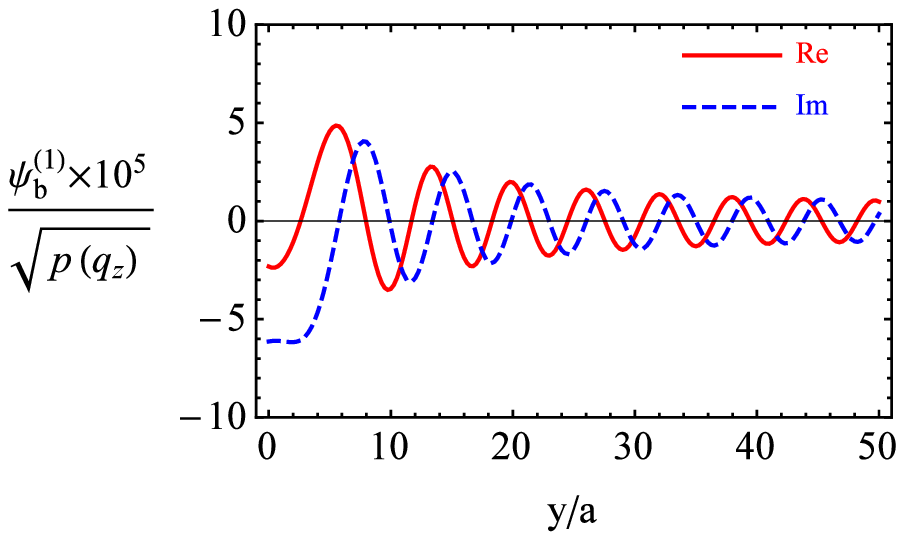}\hfill
\includegraphics[width=0.32\textwidth]{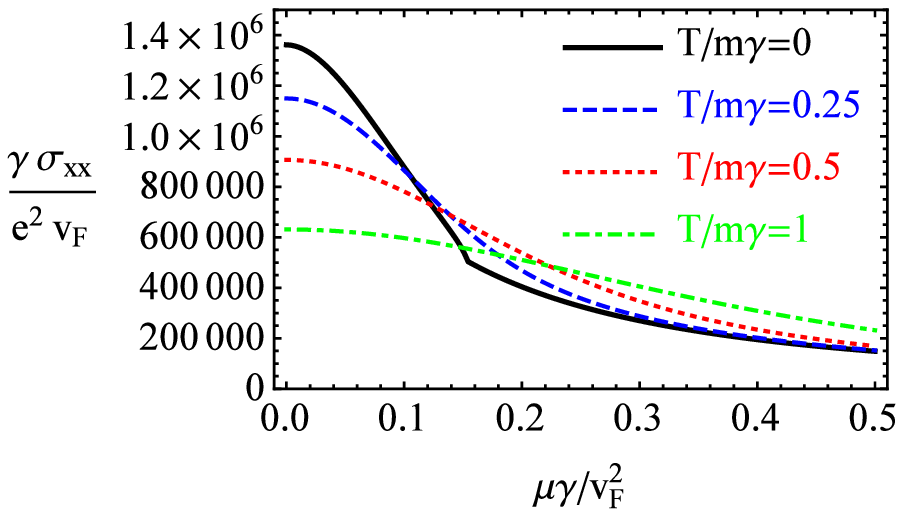}
\caption{The upper components of the surface part of scattered wave $\psi_s^{(1)}$ (panel a) and $\psi_b^{(1)}$ (panel b) as functions of $y/a$ (which quantifies the propagation into the bulk of the semimetal).
The normalized surface conductivity as a function of the electric chemical potential $\mu$ at several different temperatures (panel c). Here $p(q_z)= \gamma (m-q_z^2)/v_F$ and $\mathbf{q}$ is the wave vector of an impurity.}
\label{fig:Fermi-Arc-scattering}
\end{center}
\end{figure}

We present the dependence of the upper components of the scattered surface and bulk wave functions on $y/a$ (which quantifies the depth of the propagation into the bulk of the semimetal) for a single impurity in Figs.~\ref{fig:Fermi-Arc-scattering}(a) and \ref{fig:Fermi-Arc-scattering}(b), respectively.
Firstly, we note that $\psi^{(1)}_s(\mathbf{r})$ falls off as a power-law function of $y$, rather than an exponential function.
While this might seems surprising (the wave function of the incident wave has an exponential dependence on $y$ like that in Eq.~(\ref{psi-surface-states})),
the power-law dependence stems from the regions where Fermi arcs merge with the bulk states. Secondly and most crucially, the scattered
bulk portion of the wave function is also nonzero. Unlike their surface counterparts, these outgoing waves are propagating into the bulk, which
is suggested by the characteristic phase shift of $\pi/2$ between the real and imaginary parts of the wave function.

The dissipative character of the Fermi arcs allows us to use the perturbative methods to study their conductivity.
The corresponding analysis was done in Ref.~\cite{Gorbar:2016aov} where we included both surface and bulk states and used the Kubo response theory.
The resulting surface conductivity is shown as a function of the normalized electric chemical potential $\mu \gamma /v_F^2$ in Fig.~\ref{fig:Fermi-Arc-scattering}(c).
As one can see, the conductivity decreases with the
increase of the electric chemical potential.
This is easily explained by the growth of the Fermi arc quasiparticle width with the density of states in the bulk and, consequently, the phase space for the scattering.
It is interesting to note that the slopes of the conductivity are different at small and large
values of $\mu \gamma /v_F^2$, showing a characteristic kink at $\mu=\gamma m$ when temperature is small.
Such a Van Hove kink can be explained by the Lifshitz transition \cite{Lifshitz}, at which two disjoined sheets
of the Fermi surface  merge into a single one.
Note that the corresponding phenomenon was observed also in Dirac semimetal
$\mathrm{Na_3Bi}$~\cite{Xu:Lifshitz}.
The temperature dependence of the Fermi arc
conductivity is nonmonotonic with a local maximum at temperature that scales approximately as the
electric chemical potential. While the conductivity at small values of $\mu$ is a decreasing function of $T$, it tends to weakly
increase at large chemical potentials. Such a nonmonotonic dependency clearly suggests a nontrivial competition
between the phase space of the bulk states and the thermal effects.

It is worth noting that the region of small chemical potentials is peculiar. Technically, this is due
to the fact that the first-order corrections
become large at small energies and the leading-order
perturbative treatment is insufficient. Instead, a nonperturbative analysis is needed.
Finally, there is an interesting question related to the dissipativity of the Fermi arcs when the electric chemical potential is exactly zero.
Naively, there are no bulk states to participate in the scattering, leading to the
blow up of the Fermi arc conductivity at $\mu \to 0$. However, the study performed in Refs.~\cite{Nandkishore:2014,Rodionov:2015}
showed that there is still a nonzero density of bulk states in a disordered
Weyl semimetal with short-range impurities even at zero electric chemical potential due to nonperturbative
rare region effects~\cite{Nandkishore:2014}. For charged impurities, the nonvanishing
bulk DOS originates from the formation of electron and hole puddles~\cite{Rodionov:2015}.
Thus, the bulk states in disordered Weyl semimetals are present even at a vanishing electric chemical potential.
This suggests that the surface and bulk states of the Weyl semimetal are always coupled leading to a finite Fermi arc conductivity.

\section{Transport in the strain-induced pseudoelectromagnetic fields}
\label{sec:strain}

\subsection{Pseudoelectromagnetic fields in Dirac and Weyl semimetals}
\label{sec:strain-origin}

The interplay of axial and vector gauge fields and their interaction with chiral fermions is a delicate issue.
While the
electromagnetic fields, described by the vector potential $\mathbf{A}$, are routinely generated,
the realization of the axial gauge fields $\mathbf{A}_5$ in the high energy physics is rather uncommon. Therefore,
it is important that Dirac and Weyl materials not only demonstrate properties similar to that of the high energy relativistic matter, but
provide means to study novel quantum effects connected with the axial fields.
For example, it is shown~\cite{Zhou:2012ix,Zubkov:2015,Cortijo:2016yph,Cortijo:2016,Pikulin:2016,Grushin-Vishwanath:2016,Liu-Pikulin:2016}
that the axial vector potential $\mathbf{A}_5$ can be indeed generated by static mechanical strains in these topologically nontrivial materials. Such a potential describes axial or pseudoelectromagnetic fields.
As in graphene, the corresponding effective gauge fields capture the corrections to the kinetic energy of
quasiparticles caused by unequal modifications of hopping parameters in a strained crystal.
Unlike ordinary electric $\mathbf{E}$ and magnetic $\mathbf{B}$ fields, their pseudoelectromagnetic
counterparts $\mathbf{E}_5$ and $\mathbf{B}_5$ interact with the fermions of opposite chirality with different sign.
Consequently, the pseudomagnetic field does not break the TR
symmetry in Weyl materials (as is expected in the case of static deformations).
The characteristic strengths of the pseudomagnetic field in Dirac and Weyl materials are much smaller
than in graphene and range from about $B_5\approx0.3~\mbox{T}$, when a static torsion is applied to
a nanowire of Cd$_3$As$_2$~\cite{Pikulin:2016}, to approximately $B_5\approx15~\mbox{T}$, when a
thin film of Cd$_3$As$_2$ is bent~\cite{Liu-Pikulin:2016}. Note that the pseudoelectromagnetic fields are also related to the magnetic texture~\cite{Nomura:2017} and give rise to the regular Hall effect.

Let us briefly review the studies of the electron transport properties of Dirac and Weyl semimetals under strain.
As is shown in Ref.~\cite{Cortijo:2016}, the application of strain in Weyl semimetals changes the chiral shift and leads to an experimentally measurable CME current decaying exponentially
with time.
In addition, it was demonstrated in Ref.~\cite{Pikulin:2016} that pseudoelectromagnetic fields give rise to new and unusual manifestations of the chiral anomaly,
which can be observed by such conventional experimental probes as electrical transport, electromagnetic field emission, and ultrasonic attenuation. By using a semiclassical approach, the authors of Ref.~\cite{Grushin-Vishwanath:2016} showed
that the pseudomagnetic field $\mathbf{B}_5$ also contributes to the conductivity as $\sigma_{zz}\sim B_5^2$. In addition, it
was argued that the surface Fermi arcs can be viewed as the zero pseudo-Landau levels at $\mathbf{B}_5 \neq \mathbf{0}$
which are localized at the boundary of the system even when strain is absent.
Similarly to the unusual surface-bulk oscillations in an external magnetic field $\mathbf{B}$ described in Subsec.~\ref{sec:Fermi-arcs-quantum-oscillations} (see also Refs.~\cite{Potter:2014,Moll:2016,Gorbar:2014qta}), the strain-induced pseudomagnetic field $\mathbf{B}_5$ was also proposed to cause periodic in $1/B_5$ oscillations of the DOS~\cite{Liu-Pikulin:2016}.
The tunability of the pseudoelectromagnetic fields and their uniqueness for the relativistic-like Dirac and Weyl semimetals provide
an exciting possibility to study the quantum field effects not easily accessible in truly relativistic matter.

\subsection{Local conservation of electric charge and the consistent chiral kinetic theory}
\label{sec:strain-consistent}

Recently, the issue of the local electric charge nonconservation in chiral media with the strain-induced pseudoelectromagnetic fields
$\mathbf{E}_5$ and $\mathbf{B}_5$ came to attention~\cite{Grushin-Vishwanath:2016,Pikulin:2016}
in the framework of the chiral kinetic theory defined in Refs.~\cite{Son,Stephanov:2012ki}. Indeed, the naive continuity equations for the
chiral and electric currents are
\begin{eqnarray}
\label{CKT-dn/dt-n5}
\frac{\partial \rho_5}{\partial  t}+\bm{\nabla}\cdot\,\mathbf{J}_5 &=& -\frac{e^3}{2\pi^2}
\Big[(\mathbf{E}\cdot\mathbf{B}) +(\mathbf{E}_{5}\cdot\mathbf{B}_{5})\Big],\\
\label{CKT-dn/dt-n}
\frac{\partial \rho}{\partial  t}+\bm{\nabla}\cdot\,\mathbf{J} &=& -\frac{e^3}{2\pi^2}
\Big[(\mathbf{E}\cdot\mathbf{B}_{5}) +(\mathbf{E}_{5}\cdot\mathbf{B})\Big],
\end{eqnarray}
where $\rho$ ($\rho_5$) is the electric (chiral) charge density and $\mathbf{J}$ ($\mathbf{J}_5$) is the electric (chiral) current density.
The first equation is related to the celebrated chiral anomaly~\cite{Adler,Bell-Jackiw} and describes the nonconservation
of the chiral charge in the presence of electromagnetic or pseudoelectromagnetic fields. From a physics viewpoint,
this nonconservation can be understood as the pumping of the chiral charge between the Weyl nodes of opposite
chirality~\cite{Nielsen}. The second equation leads to a striking anomalous local nonconservation of the electric charge
when both electromagnetic and pseudoelectromagnetic fields are present. In this connection,
it was suggested~\cite{Grushin-Vishwanath:2016,Pikulin:2016}
that only the global charge conservation is respected and the local nonconservation of the electric charge describes its pumping between
the bulk and the boundary of the system.
However, this nonconservation is a very serious problem that implies the possibility to literarily create the electric charge out of nothing.

In the quantum field framework, where the same difficulty was previously
encountered too, it was proposed to correct the definition of the electric current by including the
Chern--Simons (CS) contributions~\cite{Landsteiner:2013sja}, which are also known as the Bardeen--Zumino polynomials~\cite{Bardeen-1,Bardeen-2}.
Such terms restore the
local conservation of the electric charge in the presence of both electromagnetic and pseudoelectromagnetic fields.

As we argue in Ref.~\cite{Gorbar:2016ygi}, the acute problem of the local electric charge nonconservation is resolved in the consistent chiral
kinetic theory. Technically, the resolution can be described by moving the terms on the right-hand side of Eq.~(\ref{CKT-dn/dt-n}) to its
left-hand side where they are rewritten in terms of the derivatives of the topological Chern--Simons current, i.e.,
\begin{equation}
 J^{\nu}_{\text{{\tiny CS}}} = -\frac{e^3}{4\pi^2} \epsilon^{\nu \delta \alpha \beta} A_{5,\delta} F_{\alpha \beta}.
\label{consistent-def-0}
\end{equation}
Here $\nu,\delta,\alpha,\beta=\overline{0,3}$ and $A_{5,\delta}=b_{\delta}/e+\tilde{A}_{5,\delta}$ is the axial-vector potential, which,
unlike the usual vector potential, is an observable quantity.
Indeed, in Weyl materials, $b_0$ and $\mathbf{b}$ correspond to the energy and momentum separations
between the Weyl nodes.
Further, $\tilde{A}_{5,\delta}$ is expressed through the deformation tensor and
describes strain-induced axial or pseudoelectromagnetic fields.
In components, Eq.~(\ref{consistent-def-0}) takes the following form:
\begin{eqnarray}
\rho_{\text{{\tiny CS}}} &=&-\frac{e^3}{2\pi^2}\,(\mathbf{A}_5\cdot\mathbf{B}),
\label{consistent-charge-density}
\\
\mathbf{J}_{\text{{\tiny CS}}} &=&-\frac{e^3}{2\pi^2}\,A_{5,0} \mathbf{B} + \frac{e^3}{2\pi^2}\,[\mathbf{A}_5\times\mathbf{E}].
\label{consistent-current-density}
\end{eqnarray}
For $\mathbf{B}_{5}$ to be nonzero, the axial field $\tilde{\mathbf{A}}_5$
should depend on coordinates.

It is worth noting also that Eqs.~(\ref{CKT-dn/dt-n5}) and (\ref{CKT-dn/dt-n}) in high energy physics are known as the covariant
anomaly relations that come from the fermionic sector of the theory in which the fermions of opposite chirality
are treated in a symmetric way. However, this is inconsistent with
the gauge symmetry. As was argued in Ref.~\cite{Landsteiner:2013sja} (see also Ref.~\cite{Landsteiner:2016}),
the correct physical current, satisfying the local conservation of the electric
charge, is the consistent one.
As is easy to check, the total (consistent) current $J^{\nu}_{\rm tot}=(\rho+\rho_{\text{{\tiny CS}}}, \mathbf{J}+\mathbf{J}_{\text{{\tiny CS}}})$~\cite{Bardeen-1,Bardeen-2,Landsteiner:2013sja,Landsteiner:2016}
is nonanomalous and satisfies the electric current conservation relation, i.e.,  $\partial_{\nu} J^{\nu}_{\rm tot}=0$.
Note that the CS currents are important even in the absence of strain-induced pseudoelectromagnetic fields. For example, in
the equilibrium state with $\mu_5=eb_0$, the first term in Eq.~(\ref{consistent-current-density}) ensures the absence of the
CME current~\cite{Franz,Basar:2014,Landsteiner:2016}.
Moreover, the second
term in $\mathbf{J}_{\text{{\tiny CS}}}$ describes the AHE in Weyl materials~\cite{Ran,Burkov:2011ene,Grushin-AHE,Goswami,Burkov-AHE:2014}, which is missing in the conventional chiral kinetic theory.
As we will show in Subsec.~\ref{sec:BZ-I}, the topological CS terms (\ref{consistent-charge-density}) and (\ref{consistent-current-density}) are necessarily present in a lattice models of Weyl semimetals~\cite{Gorbar:2017wpi}.

The consistent chiral kinetic theory was applied by us in order to study various collective excitations in Weyl
semimetals~\cite{Gorbar:2016sey,Gorbar:2016vvg}. We showed~\cite{Gorbar:2016sey} that in a system with a nonzero electric (chiral) chemical potential,
the background magnetic (pseudomagnetic) fields not only modify the values of the chiral magnetic and pseudomagnetic plasmon
frequencies in the long-wavelength limit, but also affect the qualitative dependence on the wave
vector. Similar modifications can be also induced by the chiral shift parameter in Weyl semimetals. The latter lifts the
degeneracy of the plasmon modes. Moreover, it strongly affects the longitudinal plasmon mode by mixing it with the
transverse ones.

Next, the helicons (i.e., transverse low-energy
gapless excitations propagating along the background magnetic field in uncompensated metals) in Weyl semimetals are also significantly altered by the chiral shift and pseudomagnetic fields.
The modification of the usual helicons by the chiral shift was predicted in Ref.~\cite{Pellegrino}.
In the absence of $B_{5}$ and at $\mu_5=eb_0=0$, the corresponding dispersion relation reads
\begin{equation}
\label{helicon-omega-app-sum-s-1}
\omega_{\text{{\tiny H}}}\big|_{B_{5}\to0, \mu_5\to0} \approx \frac{eB_0\pi v_F^2 k^2}{\pi\Omega_e^2\mu +2e^4 v_F^2 (\mathbf{B}\cdot \mathbf{b})},
\end{equation}
where the Langmuir (plasma) frequency is
\begin{equation}
\Omega_e \equiv \sqrt{\frac{4e^2}{3\pi v_F}\left(\mu^2+\mu_5^2 +\frac{\pi^2 T^2}{3}\right)}.
\label{helicon-Langmuir-def}
\end{equation}
As is easy to see, the chiral shift modifies the effective helicon mass (i.e., the denominator in Eq.~(\ref{helicon-omega-app-sum-s-1})).
Further, in the presence of $\mathbf{B}_5$, there exist the pseudomagnetic helicons~\cite{Gorbar:2016vvg} with the
energy dispersion
\begin{equation}
\label{helicon-omega-app-sum-s-2}
\omega_{\text{{\tiny PH}}}\big|_{B_{0}\to0, \mu\to0} \approx  \frac{eB_{5}\pi v_F^2 k^2}{\pi \Omega_e^2\mu_5 +2e^4 v_F^2 (\mathbf{B}_5\cdot \mathbf{b})}.
\end{equation}
Unlike the usual helicons given in Eq.~(\ref{helicon-omega-app-sum-s-1}), the pseudomagnetic ones
exist even in the absence of the background magnetic field. The necessary ingredients for the propagation of these
helicons are the strain-induced pseudomagnetic field $\mathbf{B}_{5}$ and the chiral chemical potential $\mu_5$.
Note that the latter appears naturally in the equilibrium state of a PI-odd Weyl material with a nonzero energy
separation $b_0$ between the Weyl nodes.

Experimentally, the unconventional collective excitations in Weyl materials could be detected via a relatively simple experimental setup.
It is based on the same idea that is used in metals and requires
measuring the amplitude of transmission of an electromagnetic wave through a Weyl or Dirac crystal as a function
of the (pseudo-)magnetic field (here the pseudomagnetic field can be quantified by a bending or torsion angle) or as a function of the frequency at a fixed field.
Since standing waves interfere inside the sample,
the resulting signal should oscillate
as the function of $B$ or $B_5$. The effects of the chiral shift parameter can be studied by changing the orientation of the
crystal and/or by applying strains along different directions.

\section{Anomalous transport in a two-band model of Weyl semimetals}
\label{sec:BZ}

As we saw in the previous section, the topological Chern--Simons or Bardeen--Zumino terms play a very important role in the physical properties of the collective excitations in Weyl semimetals.
In order to investigate the origin of these terms in the consistent chiral kinetic theory
as well as to study various types of transport, we consider a simple lattice model of
a Weyl semimetal~\cite{Gorbar:2017wpi,Gorbar:2017yjn,Gorbar:2017vwm}. This two-band model is described by the following Hamiltonian, which
is a generalization of the low-energy Hamiltonian given in Refs.~\cite{Volovik:1988,Fang-Bernevig:2012,Li-Roy-Das-Sarma:2016}:
\begin{equation}
\label{model-H-def}
\mathcal{H}_{\rm latt}(\mathbf{k}) = d_0(\mathbf{k}) +
\sum_{i=1}^3\sigma_i d_i(\mathbf{k}).
\end{equation}
Here
functions $d_0(\mathbf{k})$ and $\mathbf{d}(\mathbf{k})$ are periodic in momentum $\mathbf{k}=\left(k_x,k_y,k_z\right)$. Henceforth, we will omit the explicit arguments of the functions $d_0$ and $\mathbf{d}$.
As is easy to check, the dispersion relation of quasiparticles described by Hamiltonian
(\ref{model-H-def}) is given by $\epsilon_{\mathbf{k}} =  d_0 \pm |\mathbf{d}|$.
While a nonzero $d_0$ introduces an asymmetry between the
valence and conduction bands, it does not affect the key topological features of the Weyl nodes and, consequently,
should not affect the main qualitative features of the transport.
Therefore, in what follows, we will ignore it.

Let us discuss now the explicit form of the function $\mathbf{d}$.
For example, in the case of Weyl semimetals with the unit topological charge $n_{\text{\tiny W}}=1$, they have the following form:
\begin{eqnarray}
\label{model-d-def-d1}
d_1 &=& \Lambda \sin{(ak_x)},\\
\label{model-d-def-d2}
d_2 &=& \Lambda \sin{(ak_y)},\\
\label{model-d-def-d3}
d_3 &=& t_0 +t_1\cos{(ak_z)} +t_2\left[\cos{(ak_x)}+\cos{(ak_y)}\right],
\end{eqnarray}
where, for simplicity, we assumed that the lattice is cubic, i.e., all lattice spacings are equal, $a_x=a_y=c=a$, and the energy parameters
$\Lambda$, $t_0$, $t_1$, and $t_2$ are material dependent. Their characteristic
values can be obtained, for example, by fitting the dispersion relations of low-energy excitations
in $\mathrm{Na_3Bi}$ or Cd$_3$As$_2$.
When the parameters are such that $|t_0+2t_2|\leq |t_1|$, the model given in Eqs.~(\ref{model-d-def-d1})--(\ref{model-d-def-d3}) has two Weyl
nodes separated in momentum space by distance $2b_z$, where the chiral shift
parameter $b_z$ is
\begin{equation}
b_z=\frac{1}{a} \arccos{\left(\frac{-t_0-2t_2}{t_1}\right)}.
\label{model-bz}
\end{equation}
It is worth noting that the same Hamiltonian (\ref{model-H-def}), albeit with different functions $d_1$ and $d_2$, is
valid for the multi-Weyl semimetals, i.e., the semimetals with the topological charges of the Weyl nodes greater than one $n_{\text{\tiny W}}>1$ (for the explicit form of these functions, see,
e.g., Ref.~\cite{Gorbar:2017vwm}). The energy spectrum of model (\ref{model-H-def}) with the function $\mathbf{d}$ given in
Eqs.~(\ref{model-d-def-d1})--(\ref{model-d-def-d3}) is shown in Fig.~\ref{fig:lattice-Kubo}(a).
As one can easily see, the energy spectrum has a characteristic ``dome" with the height defined by $\epsilon_{0}$ (which is the value of energy $\epsilon_{\mathbf{k}}$ at $\mathbf{k=0}$).

In the subsections below, we will study the electric, chiral, and thermoelectric transport in (multi-)Weyl semimetals.

\subsection{Origin of the electric Chern--Simons currents}
\label{sec:BZ-I}

Following Ref.~\cite{Gorbar:2017wpi}, let us start from a linear
electromagnetic response in the two-band model of Weyl semimetals specified above.
The electrons interact with the electromagnetic field
through the standard interaction term $\mathcal{H}_{\rm int} = \mathbf{J}\cdot\mathbf{A}$, where the electric current density operator in the
momentum space is given by $\mathbf{J}(\mathbf{k}) = e\bm{\nabla}_{\mathbf{k}} \mathcal{H}_{\rm latt}$. Henceforth, we assume that the magnetic
field points in the $+z$ direction and is described by the vector potential in the Landau gauge $\mathbf{A}=\left(0,xB,0\right)$.

Let us first calculate the electric charge density in the background magnetic field. In the case of the vanishing electric chemical potential and temperature, the final result reads
\begin{equation}
\rho = -e^2 \int \frac{d^3\mathbf{k}}{(2\pi)^3}\, \left(\mathbf{B}\cdot\bm{\Omega}\right),
\label{lattice-topology-rho-1}
\end{equation}
where $\hat{\mathbf{d}}\equiv\mathbf{d}/|\mathbf{d}|$ and $\Omega_{z}$ denotes the $z$ (i.e., parallel to $\mathbf{B}$) component
of the Berry curvature \cite{Berry:1984,Haldane}
\begin{equation}
\Omega_{i}=\sum_{l,m=x,y,z}\frac{\epsilon_{i l m}}{4} \left(\hat{\mathbf{d}}\cdot\Big[(\partial_{k_{l}}\hat{\mathbf{d}})
\times(\partial_{k_{m}}\hat{\mathbf{d}})\Big]\right).
\label{lattice-topology-inv-Berry}
\end{equation}
In order to make clear the topological nature of the charge density (\ref{lattice-topology-rho-1}), we note that
each component of the Berry curvature can be also viewed as the Jacobian of the mapping of a two-dimensional section
of the Brillouin zone onto the unit sphere.
When integrated over the area of the cross section (in the case under consideration, this is the $k_x$-$k_y$ plane), it counts the winding
number of the mapping or the Chern number~\cite{Bernevig:2013}
\begin{equation}
\mathcal{C}(k_i)= \sum_{l,m=x,y,z}\frac{\epsilon_{i l m}}{2} \int \frac{dk_l\,dk_m}{4\pi}\, \left(\hat{\mathbf{d}}\cdot\Big[(\partial_{k_l}
\hat{\mathbf{d}})
\times(\partial_{k_m}\hat{\mathbf{d}})\Big]\right).
\label{lattice-topology-inv-1}
\end{equation}
We checked that $\mathcal{C}(k_z)$ is nonzero only for $|k_z| \leq b_z$ and after the integration over $k_z$ leads to the topological
CS expression for the electric charge density induced
by a magnetic field (see Eq.~(\ref{consistent-charge-density}) with $\mathbf{A}_5\to\mathbf{b}/e$), albeit generalized to the case of a multi-Weyl semimetal, i.e.,
\begin{equation}
\rho_{\text{{\tiny CS}}}= -e^2 B\int \frac{dk_z}{4\pi^2} \mathcal{C}(k_z) =-n_{\text{\tiny W}}\frac{e^2 (\mathbf{B}\cdot\mathbf{b})}{2\pi^2}.
\label{consistent-charge-density-BZ}
\end{equation}
Equation (\ref{consistent-charge-density-BZ}) clearly demonstrates the topological nature of the CS charge density related
to the winding number in lattice models of multi-Weyl semimetals with finite Brillouin zones.
As for the electric current density in the static background magnetic field, we found that it vanishes in the linear order in $B$. Therefore, as
expected, the CME current is absent in the equilibrium state~\cite{Franz,Basar:2014,Landsteiner:2016}.

The topological result in Eq.~(\ref{consistent-charge-density-BZ})
cannot be captured by the conventional chiral kinetic theory~\cite{Son,Stephanov:2012ki}.
This follows from the fact that the corresponding kinetic equations do not contain the chiral
shift parameter $\mathbf{b}$ at all. As we noted in Ref.~\cite{Gorbar:2017wpi}, the topological CS terms appear to be the
only contribution that the chiral kinetic theory fails to reproduce.

\begin{figure}[t]
\hspace{-0.32\textwidth}(a)\hspace{0.32\textwidth}(b)\hspace{0.32\textwidth}(c)\\[0pt]
\begin{center}
\includegraphics[width=0.32\textwidth]{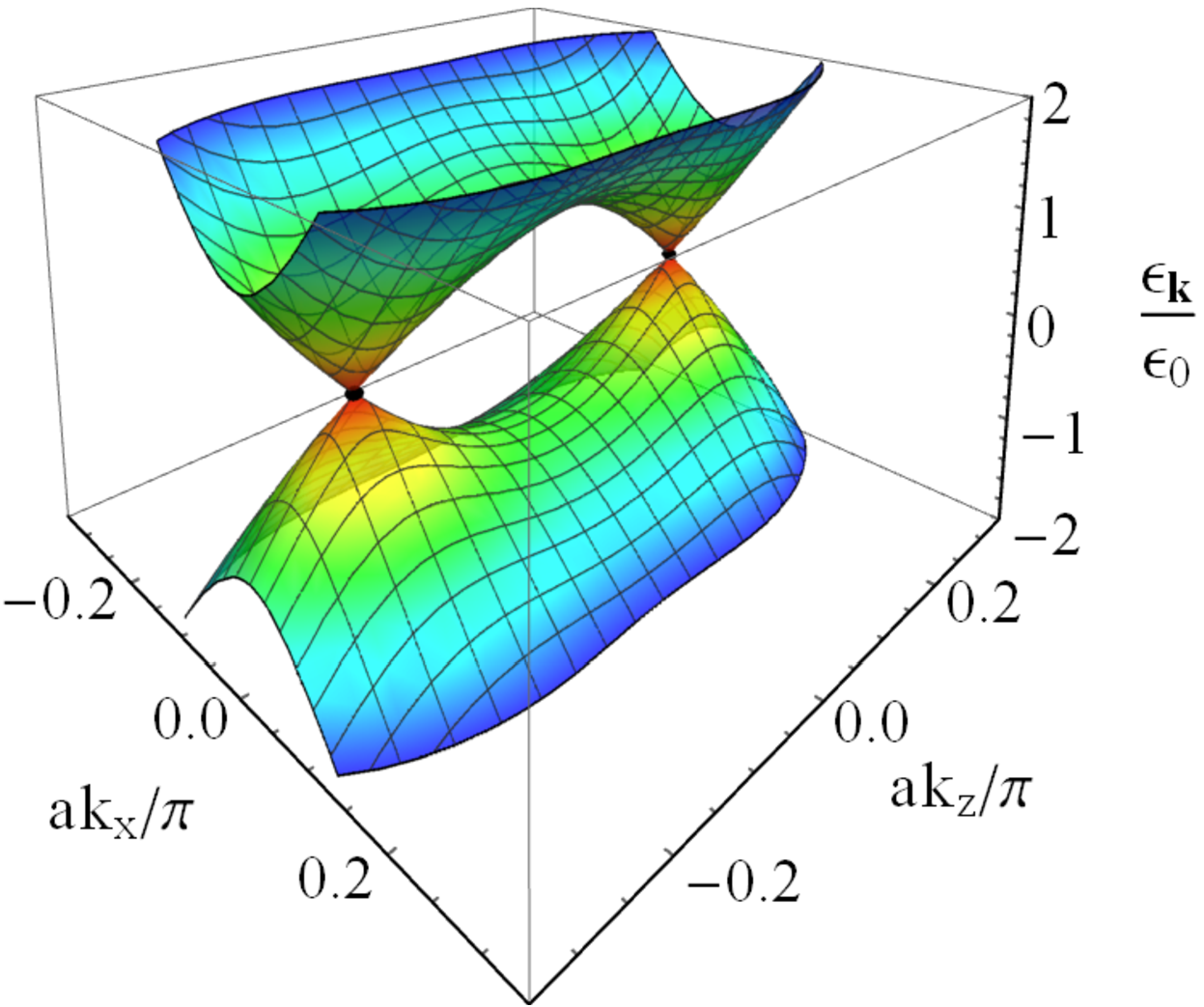}\hfill
\includegraphics[width=0.32\textwidth]{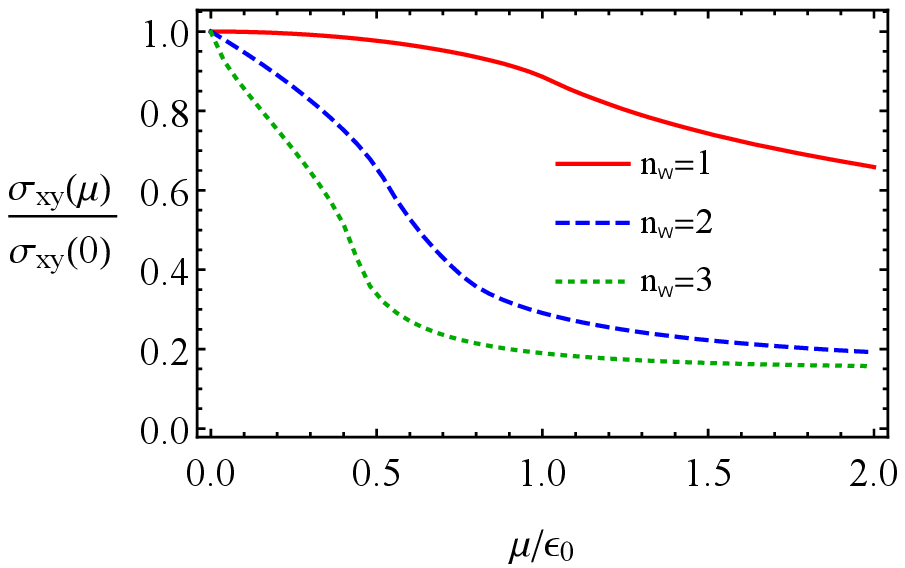}\hfill
\includegraphics[width=0.32\textwidth]{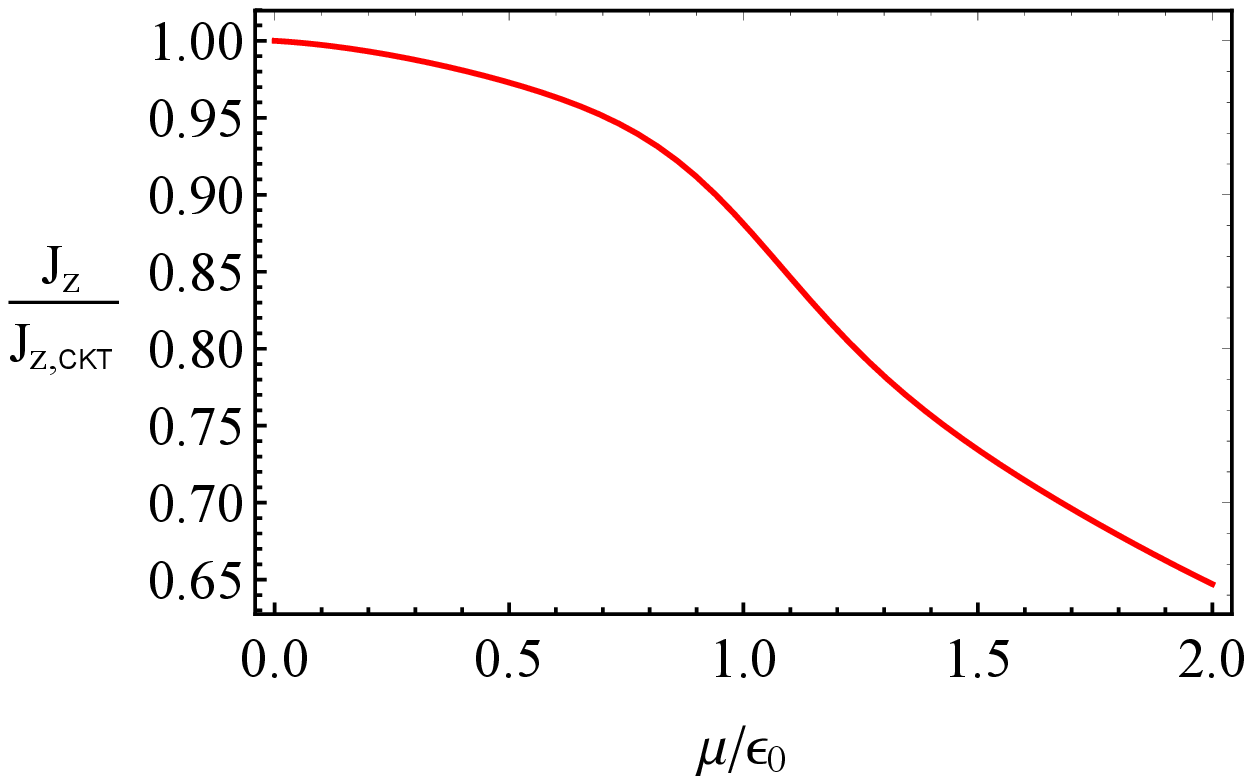}
\caption{Panel a: the energy spectrum of the two-band model (\ref{model-H-def}) with the function $\mathbf{d}$ given by Eqs.~(\ref{model-d-def-d1})--(\ref{model-d-def-d3}).
Panel b: the dependence of the zero-temperature Hall conductivity $\sigma_{xy}$ in a Weyl semimetal (red solid line), a double-Weyl semimetal (blue dashed line), and a triple- Weyl semimetal (green dotted line) on
the electric chemical potential. Panel c: the electric current density $J_z$ in the direction of the
pseudomagnetic field as a function of $\mu/\epsilon_0$.
Here $\epsilon_0$ is the value of quasiparticle energy at $\mathbf{k}=\mathbf{0}$.}
\label{fig:lattice-Kubo}
\end{center}
\end{figure}

In order to identify the topological contributions to the electric current density, one needs to
study the response of the two-band Weyl model to a background electric field. We performed the corresponding analysis in
Ref.~\cite{Gorbar:2017wpi} using the Kubo linear response theory. In the clean limit and at $T=0$, the off-diagonal conductivity tensor reads
\begin{equation}
\sigma_{nm} = -\frac{e^2}{2}\int\frac{d^3\mathbf{k}}{(2\pi)^3}
 \left(\hat{\mathbf{d}}\cdot\left[(\partial_{k_n}\hat{\mathbf{d}})\times(\partial_{k_m}\hat{\mathbf{d}})\right]\right)
 \left[1-\theta(|\mu|-|\mathbf{d}|)
 \right],
\label{Kubo-conductivity-calc-Im-Gamma0}
\end{equation}
whose only nonvanishing components are $\sigma_{xy}=-\sigma_{yx}$.
The topological nature of the off-diagonal conductivity $\sigma_{xy}$ in
Eq.~(\ref{Kubo-conductivity-calc-Im-Gamma0}) at $\mu\to 0$ is evident from the fact that it is determined by the integral of the Chern number
(\ref{lattice-topology-inv-1}) or, equivalently, the winding number. Note that the latter is equal to the topological charge of the
Weyl nodes $n_{\text{\tiny W}}$. After calculating the corresponding integral, we derive the following explicit result for the anomalous Hall conductivity:
\begin{equation}
\sigma_{xy}=-\sigma_{yx} = -n_{\text{\tiny W}}\frac{e^2 b_z}{2\pi^2},
\label{Kubo-conductivity-12}
\end{equation}
where $\mu=0$.
This result corresponds to the expected topological CS current that describes
the AHE conductivity~\cite{Ran,Burkov:2011ene,Grushin-AHE,Goswami,Burkov-AHE:2014}. Note also that it agrees with the conductivity in
Eq.~(\ref{anomaly-contribution}), albeit the magnetic field is absent now.
The dependence of the zero-temperature Hall
conductivity $\sigma_{xy}$ given in Eq.~(\ref{Kubo-conductivity-calc-Im-Gamma0}) on the electric chemical potential $\mu$ is plotted in
Fig.~\ref{fig:lattice-Kubo}(b). Due to the matter contribution, the total Hall conductivity decreases as the absolute
value of the chemical potential increases. In addition, the corresponding dependence is much steeper in the case of double- ($n_{\text{\tiny W}}=2$) and
triple-Weyl ($n_{\text{\tiny W}}=3$) semimetals.

There is another point regarding Eq.~(\ref{Kubo-conductivity-calc-Im-Gamma0}), which is worth discussing.
According to the conventional wisdom of the Fermi liquid theory, the conductivity is related only
to the states on the Fermi surface. However,
Eq.~(\ref{Kubo-conductivity-calc-Im-Gamma0}) contains the integration over all filled
quasiparticle states. Such a contradiction was resolved in Ref.~\cite{Haldane} by noting that the corresponding term can be rewritten as a Fermi
surface integral.

It is very interesting also to investigate the response of Weyl semimetals to the strain-induced pseudoelectromagnetic fields $\mathbf{B}_5$ and $\mathbf{E}_5$, which we discussed in
Subsec.~\ref{sec:strain-origin}.
According to Ref.~\cite{Pikulin:2016}, strains in Weyl materials lead to the following additional
terms in the Hamiltonian:
\begin{equation}
\label{lattice-B5-dh}
\delta h_{\rm strain}= \Lambda \left(u_{xz}\sigma_x +u_{yz}\sigma_y\right)\sin{(ak_z)}
-t_1 u_{zz} \sigma_z \cos{(ak_z)},
\end{equation}
where $u_{nm}=\left(\partial_{n}u_m+\partial_{n}u_m\right)/2$ is the symmetrized strain tensor and
$\mathbf{u}=(u_x,u_y,u_z)$ is the displacement vector. The notion of the strain-induced axial potential $\mathbf{A}_5$
naturally arises, because, in the vicinity of Weyl nodes, $\delta h_{\rm strain}$ can be interpreted as the interaction
Hamiltonian of Weyl quasiparticles with the background axial gauge field
\begin{equation}
\label{lattice-B5-Franz-gauge}
\mathbf{A}_5= -\frac{1}{ea}\left[u_{xz}\sin{(ab_z)},u_{yz}\sin{(ab_z)},u_{zz}\cot{(ab_z)}\right].
\end{equation}
For example, the strain-induced pseudomagnetic field
along the $z$ direction can be induced by applying torsion to a wire
made of a Weyl semimetal~\cite{Pikulin:2016}, i.e.,
$\mathbf{B}_5\equiv \bm{\nabla}\times\mathbf{A}_5=\frac{\theta}{L e a} \sin{(ab_z)} \hat{\mathbf{z}}$,
where $\mathbf{u}= \theta z [\mathbf{r}\times\hat{\mathbf{z}}]/L$, $\theta$ is the torsion angle, and $L$ is the length of the crystal.

Our numerical calculations in Ref.~\cite{Gorbar:2017wpi} show that $\rho=J_{x}=J_y=0$ if
only the pseudomagnetic field is present. Therefore, unlike the magnetic field, the pseudomagnetic one
does not induce any electric charge density. On the other hand, the component of the electric current along the direction of the pseudomagnetic
field is nonzero and is shown as a function of $\mu$ in Fig.~\ref{fig:lattice-Kubo}(c). At sufficiently small values of $\mu$,
the electric current agrees with the corresponding expression in the chiral kinetic theory (CKT)~\cite{Zhou:2012ix,Grushin-Vishwanath:2016}
\begin{equation}
J_{z,\text{{\tiny CKT}}}= -\frac{e^2 \mu B_5}{2\pi^2}.
\label{lattice-B5-Jz-top}
\end{equation}
However, as is obvious from Fig.~\ref{fig:lattice-Kubo}(c), the calculated current
deviates from Eq.~(\ref{lattice-B5-Jz-top}) as $\mu$ increases.
Physically, this is related to the fact that the interpretation of the strain-induced background field (\ref{lattice-B5-Franz-gauge}) as
a conventional axial vector potential $\mathbf{A}_5$ deteriorates outside the immediate
vicinity of the Weyl nodes. In addition, we note that the strain-induced pseudoelectric field $\mathbf{E}_5$ leads
to a sort of the dynamical piezoelectric effect with
$\rho\propto E_{5}$, which is, however, not topological.

\subsection{Chiral response}
\label{sec:BZ-II}

In this subsection, we present the analytical results for the chiral (or, equivalently, valley) charge and current densities in background
electromagnetic and strain-induced pseudoelectromagnetic fields.
The corresponding study in Ref.~\cite{Gorbar:2017yjn} is motivated by the fact that, as was argued in Refs.~\cite{Landsteiner:2013sja,Landsteiner:2016}, in addition to the CS terms
in the electric charge and current densities, their chiral or axial counterparts should be also accounted for.
In the four-vector notation, the corresponding chiral CS current is given by
$J^{\nu}_{5,\, \text{{\tiny  CS}}} =  -e^2\epsilon^{\nu \delta \alpha \beta} A_{5,\delta} F_{5,\alpha \beta}/(12\pi^2)$.

Before discussing the chiral charge and current densities in the two-band model of Weyl semimetals,
it is necessary to note
that while the concept of chirality is well-defined only for the quasiparticles in the vicinity of the Weyl
nodes, its generalization to the whole Brillouin zone is problematic. Therefore, we consider two definitions
of chirality. The first definition is the standard one for chirality in systems with a linear dispersion law
$\mathcal{H}\sim\sum_{i=x,y,z}v_ik_i\sigma_i$ generalized to the entire Brillouin zone, i.e.,
\begin{equation}
\chi_1(\mathbf{k}) \equiv \sign{v_xv_yv_z},
\label{lattice-B5-chi-1}
\end{equation}
where $v_i\equiv \partial_{k_i}d_i$ is the quasiparticle velocity.
The second definition is specific for the two-band model given by Eqs.~(\ref{model-H-def})--(\ref{model-d-def-d3}) and is related to
the reflection symmetry $k_z \to -k_z$ of the Hamiltonian, i.e.,
\begin{equation}
\chi_2(\mathbf{k})  \equiv -\sign{k_z}.
\label{lattice-B5-chi-3}
\end{equation}
Due to the existence of two valleys in the model at hand (which is directly related to the separation of the Weyl nodes of opposite chirality),
such a symmetry can be also identified as a valley symmetry and makes sense even for the states away from the Weyl nodes.
Note, however, that definition (\ref{lattice-B5-chi-3}) is limited only to Weyl materials with a broken TR symmetry.
While in the vicinity of the Weyl nodes both definitions of chirality in Eqs.~(\ref{lattice-B5-chi-1}) and (\ref{lattice-B5-chi-3}) are
completely equivalent, they differ for the states far from the Weyl nodes.

Similarly to Subsec.~\ref{sec:BZ-I}, we begin with the response to the background magnetic field. The
topological or vacuum part of the chiral charge density $\rho^{5}$ at
$\mu=T=0$ reads
\begin{equation}
\rho^{5}_0 = -e^2 \int \frac{d^3\mathbf{k}}{(2\pi)^3}\, \chi(\mathbf{k})
\left(\mathbf{B}\cdot\bm{\Omega}\right),
\label{lattice-topology-rho-5-1}
\end{equation}
where $\chi(\mathbf{k})$ is a chirality function that is given either by Eq.~(\ref{lattice-B5-chi-1}) or (\ref{lattice-B5-chi-3}).

Because of the additional factor $\chi(\mathbf{k})$, the chiral charge density does
not have the same topological robustness as the electric charge density (\ref{lattice-topology-rho-1}). Nevertheless, it may be convenient
to define a chiral analog of the Chern number (\ref{lattice-topology-inv-1}),
\begin{equation}
\mathcal{C}_{\chi}(k_i)= \sum_{l,m=x,y,z} \frac{\epsilon_{i l m}}{2} \int \frac{dk_l\,dk_m}{4\pi}\, \chi(\mathbf{k})\, \left(\hat{\mathbf{d}}\cdot
\Big[(\partial_{k_l}\hat{\mathbf{d}})
\times(\partial_{k_m}\hat{\mathbf{d}})\Big]\right).
\label{lattice-first-C2}
\end{equation}
For the chirality defined by Eq.~(\ref{lattice-B5-chi-3}), there is a simple
relation between the chiral and standard Chern numbers: $\mathcal{C}_{\chi_2}(k_z) = -\sign{k_z}\mathcal{C}(k_z)$.
Obviously, $\mathcal{C}_{\chi_2}$ takes
only integer values and is zero for $|k_z|>b_z$. Taking into account the reflection symmetry $k_z \to -k_z$ of the model, $\mathcal{C}_{\chi_2}$
can be considered as a symmetry-protected topological invariant.
In contrast, $\mathcal{C}_{\chi_1}$ is generically noninteger and depends on the details of the model.
We showed~\cite{Gorbar:2017yjn} that the ``vacuum" part of the chiral charge vanishes $\rho_{0}^{5}=0$, because the chiral
Chern number is an odd function of $k_z$.
Therefore, no chiral charge is induced by the magnetic field.

After integrating over the Brillouin zone, we find that only the longitudinal component
(with respect to $\mathbf{B}$) of the chiral current density is nonzero. For sufficiently small values of the
electric chemical potential, i.e., $|\mu|<\epsilon_0$, when two separate chiral sheets
of the Fermi surface are formed, the chiral current density coincides with the well-known CSE expression in linearized
effective models~\cite{Vilenkin:1980fu,Metlitski:2005pr,Newman:2005as}, i.e.,
$\mathbf{J}_{\text{{\tiny CSE}}}^5 = -e^2\mathbf{B}\mu/(2\pi^2)$.
However, the dependence of the chiral current density changes when the system undergoes the Lifshitz transition at $|\mu| = \epsilon_0$
(see also Fig.~\ref{fig:lattice-Kubo}(a)).
Indeed, for chemical
potentials larger than $\epsilon_0$, the chirality becomes ill-defined and, as a consequence,
the chiral current deteriorates compared to that in the linearized models.
It is interesting to note that the results are almost the same for both definitions of chirality.
This follows from the fact that the current for small
values of $\mu$ is determined by the states in the vicinity of Weyl nodes where
$\chi_1(\mathbf{k}) \simeq \chi_2(\mathbf{k})$.
Further, integrating over the Brillouin zone, we found~\cite{Gorbar:2017yjn} a nonzero chiral charge density induced by the background electric
field. It depends on the disorder strength and, consequently, is not topologically protected.

For completeness, we studied also the chiral response to the pseudoelectromagnetic fields.
The analysis of the chiral charge density in the pseudomagnetic fields reveals, however, that the actual chiral charge density differs from the
expected topological result $\rho_{\text{{\tiny CS}}}^5= -e^2B_5 b_z/(6\pi^2)$~\cite{Landsteiner:2013sja,Landsteiner:2016}. While
the inability to reproduce the chiral CS terms may seem surprising,
this might have been expected taking into account the subtlety with the implementation of the chirality in lattice models
as well as the effective character of the pseudoelectromagnetic fields. Indeed, the latter are not coupled minimally in the whole Brillouin
zone. The same is true for the response of the chiral current density to the pseudoelectric field, where the relativistic quantum field theory result
$\sigma_{5,\rm \text{{\tiny AHE}}} = -e^2b_z/(6\pi^2)$ is not reproduced either.

These findings are in a strong contrast with the results for the electric CS terms studied in the
previous subsection. In essence, the key to understanding the distinction between
these two sets of observables is
the conceptual difference between the definitions of the exactly conserved electric charge and a much less unambiguous concept of
chirality on a lattice. Therefore, the chiral analogs of the electric Chern--Simons terms in Weyl materials
do not possess a mathematically well-defined topological status.

However, this does not diminish the potential practical value of the chiral or, equivalently, valley transport.
Indeed, even a nontopological anomalous chiral Hall effect could find useful applications,
which rely on the chirality or valley degrees of freedom.
Therefore, let us briefly discuss the physical meaning of the chiral or valley current.
In contrast to the electric current, the chiral one is not directly observable.
However, it still plays an important role in physical processes in Weyl semimetals.
For example, it is important for the collective excitations. An interplay
between the electric and chiral currents allows for a unusual type of collective excitations known as the chiral
magnetic wave~\cite{Kharzeev-Yee:2011}. Such a wave is a self-sustained mode in which
the chiral current induces a fluctuation of the chiral chemical potential that drives the electric current via
the CME. On the other hand, the electric current produces a fluctuation of the electric chemical potential.
The latter leads to the chiral current via the CSE that closes the cycle.
In addition, the chiral current also affects the properties of the chiral plasmons in
a qualitative way~\cite{Gorbar:2016ygi,Gorbar:2016sey}.
Therefore, the observation of various collective modes
could provide an additional indirect observation of the chiral current.

\subsection{Anomalous thermoelectric phenomena}
\label{sec:BZ-III}

In this subsection, we complete the investigation of the transport phenomena in the two-band model of multi-Weyl semimetals by studying
the thermoelectric response. In literature, the thermal transport in Dirac and Weyl semimetals was investigated in
Refs.~\cite{Kim:2014,Lundgren:2014hra,Spivak:2016,Sharma:2016-Weyl}
by using a semiclassical approach of the chiral kinetic theory. It was shown that the chiral anomaly
plays an important role producing the characteristic quadratic dependence of the thermal conductivity
on the magnetic field when the temperature gradient is parallel to the field.
In addition, the Wiedemann--Franz law does not hold when an external magnetic field is present~\cite{Kim:2014}.
Such an effect was suggested to be another hallmark of Weyl semimetals that originates from their nontrivial topology.
Further, by using the chiral kinetic theory with the Berry curvature obtained in a two-band model, it was shown~\cite{Sharma:2016-Weyl} that, in addition to
the conventional Nernst effect in a magnetic field, which was recently measured in NbP~\cite{McCormick:2017}, there is also an anomalous Nernst response, which is determined by the chiral shift.
Comparing to the Weyl semimetals with the unit topological charge $n_{\text{\tiny W}}=1$, the anomalous thermoelectric transport coefficients in double-Weyl ones contain the topological charge multiplier $n_{\text{\tiny W}}=2$~\cite{Chen-Fiete:2016}
In addition, the transport in the latter
exhibits an interesting directional dependence.

Phenomenologically, the electric and heat transport current densities in
terms of a background electric field and a temperature gradient (see, e.g., Ref.~\cite{Mahan-book}) are given by
\begin{eqnarray}
\label{Kubo-Je-Jq-gen-be}
J_n &=& e^2 L_{nm}^{11} E_{m} -eL_{nm}^{12}\nabla_m\left(\frac{1}{T}\right), \\
\label{Kubo-Je-Jq-gen-ee}
J_n^Q &=& -e\frac{1}{T}L_{nm}^{21}E_{m} +L_{nm}^{22}\nabla_m\left(\frac{1}{T}\right),
\end{eqnarray}
where $n$ and $m$ are the spatial indices (i.e., $x$, $y$, $z$) and the thermodynamic forces are
defined to comply with the Onsager reciprocal relation $L_{nm}^{12}=L_{mn}^{21}$.
As is obvious from Eq.~(\ref{Kubo-Je-Jq-gen-be}), the transport coefficients $L_{nm}^{11}$ and
$L_{nm}^{12}$ define the electric current densities induced by a background electric field and
a temperature gradient, respectively. The coefficient $L_{nm}^{11}$ is directly related
to the electric conductivity tensor as $L_{nm}^{11}\equiv \sigma_{nm}/e^2$.
From Eq.~(\ref{Kubo-Je-Jq-gen-ee}), we see that $L_{nm}^{21}$ and $L_{nm}^{22}$ define the heat current
density in response to an electric field and temperature gradient, respectively.

Using the Kubo linear response theory, we calculated the corresponding transport coefficients in Ref.~\cite{Gorbar:2017vwm}.
However, it is worth noting that the standard Kubo formalism is unable to capture the thermoelectric
coefficients $L_{nm}^{12}$, $L_{nm}^{21}$, and $L_{nm}^{22}$ correctly in a general case. In
particular, it may fail in the presence of nonzero gradients of the electric chemical potential
and/or temperature~\cite{Luttinger,Smrcka:1977}. The problem is related to the thermodynamic nature of the corresponding
driving forces, which cannot be captured by an interaction Hamiltonian alone without a simultaneous
adjustment of local thermodynamic equilibrium.
By following the Luttinger's
approach~\cite{Luttinger}, it was shown in Refs.~\cite{Cooper-Ruzin:1997,Qin-Niu:2011} that, in addition to the standard Kubo coefficients,
there are also terms in the local currents that are related to the electromagnetic orbital magnetization
$\mathbf{M}$ and the so-called heat magnetization $\mathbf{M}^Q=\mathbf{M}^E+\mu\mathbf{M}/e$. Here $\mathbf{M}^E$ corresponds to
the gravitomagnetic energy. These magnetizations sustain additional transport currents that are
proportional to the thermodynamic forces and the local magnetization~\cite{Cooper-Ruzin:1997,Qin-Niu:2011}. Such currents are necessary for the
correct description of the thermoelectric response as well as for reproducing the Onsager reciprocal relations.

\begin{figure}[t]
\begin{center}
\hspace{-0.32\textwidth}(a)\hspace{0.32\textwidth}(b)\hspace{0.32\textwidth}(c)\\[0pt]
\includegraphics[width=0.32\textwidth]{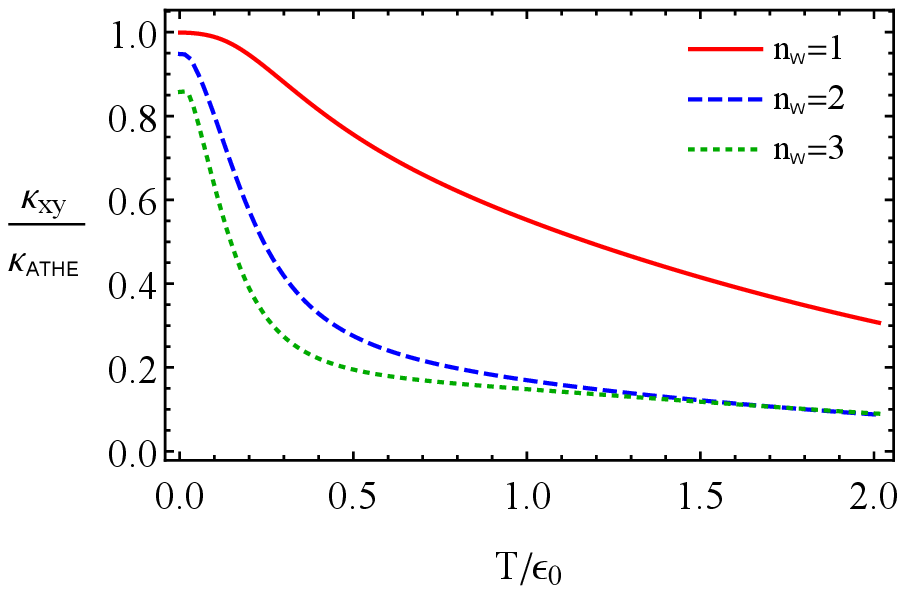}\hfill
\includegraphics[width=0.32\textwidth]{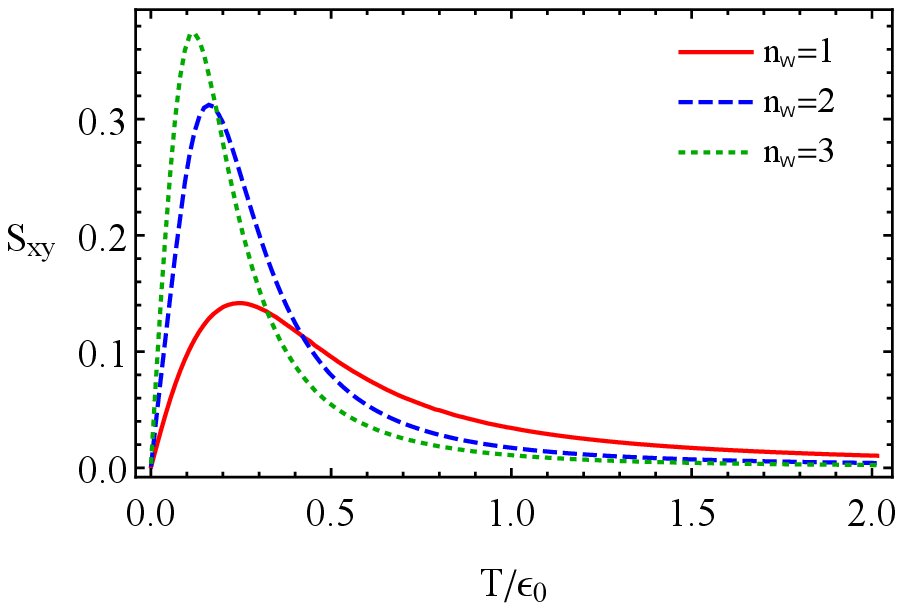}\hfill
\includegraphics[width=0.32\textwidth]{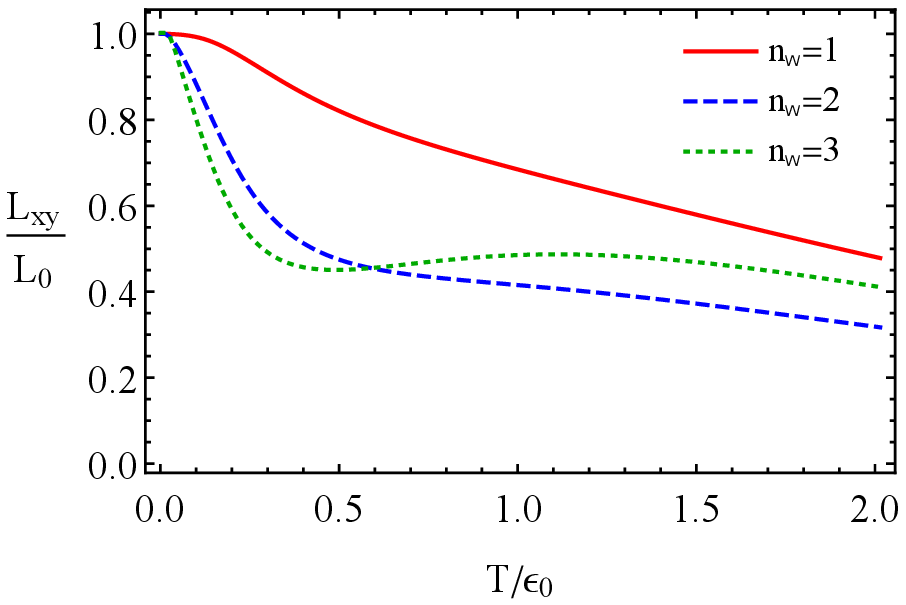}
\caption{The dependence of the off-diagonal components of the thermal conductivity $\kappa_{xy}$ (panel a), thermopower $S_{xy}$ (panel b), and relative Lorenz number $L_{xy}/L_0=\kappa_{xy}/(e^2L_0  T L_{xy}^{11})$ (panel c) on
temperature in a Weyl semimetal (red solid line), a double-Weyl semimetal (blue dashed line), a triple-Weyl
semimetal (green dotted line) at fixed $\mu=0.1\,\epsilon_0$. The results are plotted in the clean limit $\Gamma=0$.}
\label{fig:Kubo-E-kappa-S-T-L}
\end{center}
\end{figure}

The explicit form of the coefficients $L_{nm}^{11}$, $L_{nm}^{12}$, $L_{nm}^{21}$, and $L_{nm}^{22}$ is given in Ref.~\cite{Gorbar:2017vwm}. Here, for the sake of brevity, we
start directly from the definition of the
experimentally relevant thermal conductivity tensor $\kappa_{nm}$.
By enforcing a setup in which there is a thermal current, but an electric current is absent (see, e.g., Ref.~\cite{Mahan-book}),
$\kappa_{nm}$ reads as
\begin{equation}
\label{Kubo-thermo-all-thermal-cond-def}
\kappa_{nm}=\frac{1}{T^2}\left[L_{nm}^{22} - \frac{1}{T}L_{nl}^{21} (L^{11})^{-1}_{lj} L_{jm}^{12}\right].
\end{equation}
We present our results for the most interesting (and potentially anomalous) off-diagonal component of the conductivity tensor
$\kappa_{xy}=-\kappa_{yx}$ in Fig.~\ref{fig:Kubo-E-kappa-S-T-L}(a) as a function of temperature.
The off-diagonal component $\kappa_{xy}$ describes the thermal Hall effect, which is also anomalous
and related to the topological charge of the Weyl nodes. In the limit $T\to0$ and $\mu\to0$, we have the following anomalous thermal Hall effect
(ATHE) conductivity:
\begin{equation}
\kappa_{xy}=-\frac{\pi^2 T }{3}\int\frac{d^3\mathbf{k}}{(2\pi)^3} \Omega_{z}(\mathbf{k}) =\kappa_{\text{{\tiny ATHE}}} =-n_{\text{\tiny W}}\frac{Tb_z}{6}.
\label{Kubo-E-ATHE}
\end{equation}
It corresponds to the heat current
\begin{equation}
\mathbf{J}^{Q}_{\text{{\tiny ATHE}}} = -\frac{n_{\text{\tiny W}} T^3}{6} \left[\bm{\nabla}\left(\frac{1}{T}\right)\times \mathbf{b}\right],
\label{Kubo-E-BZ-thermal}
\end{equation}
which plays a principal role in reproducing the Wiedemann--Franz law.
One can easily see in Fig.~\ref{fig:Kubo-E-kappa-S-T-L}(a) that the thermal Hall conductivity generally deteriorates with $T$.
While $\mathbf{J}^{Q}_{\text{{\tiny ATHE}}}$ clearly resembles the CS
term in the electric current, it is induced by thermally excited
quasiparticles and, therefore, cannot be completely identified as an analog of the topological CS current.

Another important characteristic of the thermal transport is the thermopower or the Seebeck tensor defined as
\begin{equation}
\label{Kubo-L-all-Seebeck-def}
S_{nm}=-\frac{1}{eT^2}(L^{11})^{-1}_{nl} L_{lm}^{12}.
\end{equation}
Physically, it measures the magnitude of an induced thermoelectric voltage in response to a temperature gradient.
We plot the dependence of $S_{xy}=-S_{yx}$ on temperature in Fig.~\ref{fig:Kubo-E-kappa-S-T-L}(b).
While all three types of Weyl semimetals share the same bell-shape dependencies
on $T$, the maximal values of the off-diagonal coefficients are considerably larger
in materials with the topological charge $n_{\text{\tiny W}} >1$.

A few words are in order here regarding the off-diagonal coefficient $L_{xy}^{21}$. From a physics viewpoint, this coefficient describes the
heat current perpendicular to the external electric field and the chiral shift, i.e.,
\begin{equation}
\label{Kubo-Lxy-21-xxx}
\mathbf{J}^{Q}_{\text{{\tiny Ett}}} = -\frac{e}{T} L_{xy}^{21} \left[\mathbf{E}\times \hat{\mathbf{b}}\right],
\end{equation}
where $\hat{\mathbf{b}} \equiv \mathbf{b}/|\mathbf{b}|$ and the
ratio $L_{xy}^{21}/T$ vanishes in the limit $\mu=T=0$.
In essence, the relation in Eq.~(\ref{Kubo-Lxy-21-xxx}) describes the inverse of the Nernst effect and is sometimes
called the Ettingshausen--Nernst effect. Further, in view of the Onsager reciprocal relation, $L_{nm}^{21}=L_{nm}^{12}$,
a similar expression is true for the thermoelectric transport coefficients in the electric current. In
particular, the Nernst conductivity is defined by $L_{xy}^{12}$ and the corresponding current reads
\begin{equation}
\label{Kubo-Lxy-12-xxx}
\mathbf{J}_{\text{{\tiny Ner}}} = -e L_{xy}^{21} \left[\bm{\nabla}\left(\frac{1}{T}\right)\times \hat{\mathbf{b}}\right].%
\end{equation}
This result is in agreement with the previous findings in Ref.~\cite{Sharma:2016-Weyl}, where the anomalous
Nernst response was predicted for the multi-Weyl semimetals. Note that another form of the anomalous Nernst response can be also induced by the conformal anomaly in a background magnetic field~\cite{Chernodub-Cortijo:2017}.

Let us discuss now the Wiedemann--Franz law, which relates the thermal and electrical conductivities. It is generally held when the same well-defined quasiparticles are responsible for both types
of conduction. In terms of the transport coefficients, the Wiedemann--Franz law reads
\begin{equation}
\label{Kubo-L-all-WFL-def}
\kappa_{nm}= e^2 L_0  T  L_{nm}^{11},
\end{equation}
where $L_0=\pi^2/(3e^2)$ is the Lorenz number.
In order to study this relation in multi-Weyl semimetals, we plot the dependence of the off-diagonal component of the relative Lorenz number $L_{xy}/L_0\equiv \kappa_{xy}/(e^2 L_0  T L_{xy}^{11})$ on temperature in Fig.~\ref{fig:Kubo-E-kappa-S-T-L}(c).
As expected, the
Wiedemann--Franz law holds in the limit of small $T$. As for the deviations at nonzero $T$, they
first quickly increase with temperature and then gradually slow down. Note that the deviations in the intermediate region of
temperatures are smaller in the $n_{\text{\tiny W}}=1$ Weyl semimetal than in the double- and triple-Weyl semimetals.
It is important to emphasize that the Wiedemann--Franz law holds exactly in the clean limit $\Gamma=0$ and $T \to 0$.
This result demonstrates that a nontrivial topology in multi-Weyl semimetals by itself does not
cause any violation of the Wiedemann--Franz law. Such a conclusion also agrees with the analysis in the linearized kinetic
theory~\cite{Lundgren:2014hra}.

As in the case of the electric response, the nontrivial
topology of the electron structure of multi-Weyl semimetals also plays a profound role in the thermoelectric transport.
Indeed, the chiral shift allows for the anomalous Nernst effect, which implies the existence of an electric
current in response to a thermal gradient even in the absence of an external magnetic field.
Similarly, the off-diagonal components of the heat current are induced by a thermal gradient and an
electric field. They describe the anomalous thermal Hall and Ettingshausen--Nernst effects, respectively.
Our calculations show that all anomalous thermoelectric coefficients, including the anomalous
Hall, Ettingshausen--Nernst, Nernst, and thermal Hall effects in multi-Weyl semimetals, contain an additional
multiplication factor, which in the limit of zero temperature and electric chemical potential, is the integer topological charge of
the Weyl nodes.

\section{Summary}
\label{sec:Summary}

In this review, we discussed different aspects of a charge transport in Dirac and Weyl semimetals subjected to
external electromagnetic and axial (or pseudo-) electromagnetic fields. A special attention was paid to the nontrivial topological
properties of these materials. In Weyl semimetals, they are primarily connected with the energy and momentum separations
of the Weyl nodes. The physical properties of Dirac semimetals are also nontrivial, however. By studying the phase diagram of a Dirac semimetal in a background magnetic field at a nonzero charge density,
we showed that there exists a critical value of the electric chemical potential at which a first-order phase transition takes
place~\cite{Gorbar:2013qsa}. For small chemical potentials, the quasiparticle spectrum is a gapped state with the dynamically generated Dirac
mass. Such a state is parity and time-reversal invariant.
The supercritical phase is a gapless state where each Dirac point splits into
two Weyl nodes of opposite chirality. The splitting is quantified by a dynamically generated
chiral shift~\cite{Gorbar:2013qsa}. The latter is directed along the magnetic field and its magnitude is determined by the quasiparticle
charge density, the strength of the magnetic field, and the strength of the interaction. It worth noting that the chiral shift is a three
dimensional analog of the Haldane gap in graphene and breaks the time-reversal symmetry.

By making use of the Kubo response theory, the bulk conductivity of Dirac and Weyl semimetals in a magnetic field was
calculated~\cite{Gorbar:2013dha}. It is found that the longitudinal (with respect to the direction of the magnetic field)
magnetoresistivity is negative, i.e., decreases with $B$, at sufficiently large magnetic fields for both types of
semimetals. Physically, this phenomenon is connected with the dimensional spatial reduction $3 \to 1$ in the dynamics of the lowest
Landau level. The off-diagonal component of the transverse conductivity in Weyl semimetals
contains an anomalous contribution that is directly proportional to the chiral shift and present even in the absence of a magnetic field.
This contribution describes the anomalous Hall effect~\cite{Ran,Burkov:2011ene,Grushin-AHE,Goswami,Burkov-AHE:2014}.
It is worth noting that in the presence of the background magnetic field this term comes
exclusively from the lowest Landau level and, as expected,
is independent of temperature, the electric chemical potential, and the strength of the field
(the situation is clearly similar to that for the chiral anomaly~\cite{Ambjorn}).
A negative longitudinal magnetoresistivity is expected also in Dirac semimetals and, thus, is
not an unambiguous signature of Weyl semimetals.
Moreover, some effects induced by the chiral shift could potentially be observed, because the latter is dynamically generated when a background magnetic field is applied~\cite{Gorbar:2013qsa}.

Further, by making use of a simple model, we discussed the surface Fermi arc states. Such states allow
for an unconventional Fermiology of the Dirac and Weyl semimetals, where the disjoined arcs on the
opposite surfaces of the sample are connected via the bulk states~\cite{Potter:2014}.
These closed magnetic orbits lead to the quantum
oscillations of the density of states periodic in the inverse magnetic field. Their surface-bulk nature is revealed in the
characteristic $T_{1/B}\propto \cos{\theta}$ dependence of the oscillations period on the angle $\theta$ between the magnetic field
and the surface normal. However, these oscillations are almost unobservable in the thick samples where the bulk states dominate.
In addition, we showed~\cite{Gorbar:2014qta} that the interaction effects can change the arc length and introduce the unconventional dependence
on the magnetic field orientation.

Due to the topologically protected nature of the Fermi arc states, one may naively expect that the surface charge transport in Weyl
semimetals should be nondissipative. However, we found~\cite{Gorbar:2016aov} that this is not correct. Contrarily, the
Fermi arc transport is dissipative. The origin of the dissipation is the scattering of the surface Fermi arc
states into the bulk of the semimetal. Nondecoupling of the surface
and bulk sectors in the low-energy theory of Weyl semimetals invalidates the usual argument
of a nondissipative transport due to the one-dimensional nature of the arc states. Therefore, the corresponding scattering rate is nonzero
and can be estimated even in a perturbative approach.
The nondecoupling implies that there is no well-defined effective theory of the Fermi arcs in disordered Weyl semimetals. Note that
while the dephasing of the Fermi arc surface states and dissipation are sustained by the scattering of quasiparticles from the surface into the
bulk and vice versa in Weyl semimetals, this is not true in the case of topological insulators, where the decoupling is protected by a gap in
the bulk.

Remarkably, as is discussed in Refs.~\cite{Zhou:2012ix,Zubkov:2015,Cortijo:2016yph,Cortijo:2016,Grushin-Vishwanath:2016,Pikulin:2016,Liu-Pikulin:2016}, the Weyl and Dirac semimetals allow for the axial or pseudoelectromagnetic fields $\mathbf{E}_5$ and $\mathbf{B}_5$. In essence, these fields interact with the fermions of opposite chirality with different sign. This leads to various effects, including the corrections to electric conductivity, electromagnetic field emission, ultrasonic attenuation, and periodic in $1/B_5$ quantum oscillations. The uniqueness of $\mathbf{E}_5$ and $\mathbf{B}_5$ allows one to study different phenomena not easily accessible in a high energy relativistic matter.
Moreover, the quasiclassical description of the charge transport in Weyl semimetals is also significantly affected by their nontrivial topology.
We showed~\cite{Gorbar:2016ygi} that the correct kinetic description can be done using the consistent chiral kinetic theory.
Such a theory
necessarily includes the Chern--Simons~\cite{Landsteiner:2013sja} or, equivalently, Bardeen--Zumino~\cite{Bardeen-1,Bardeen-2} contributions to the electric charge and current densities.
This makes the theory consistent with the local conservation
of the electric charge in electromagnetic and strain-induced pseudoelectromagnetic fields.
Moreover, these terms play an important role even in the absence of pseudoelectromagnetic fields. The Chern--Simons terms in the consistent chiral kinetic theory allow one to correctly
describe the anomalous Hall effect in Weyl materials~\cite{Ran,Burkov:2011ene,Grushin-AHE,Goswami,Burkov-AHE:2014} and to reproduce the vanishing CME current in
the equilibrium state of chiral plasma~\cite{Franz,Basar:2014,Landsteiner:2016}. In addition,
the topological terms affect the spectra of various collective modes, including the chiral (pseudo-)magnetic plasmons~\cite{Gorbar:2016sey} and
helicons~\cite{Pellegrino,Gorbar:2016vvg}.
The anomalous features of these modes could be used to identify them in experiments and suggest the
efficient practical means of extracting the chiral shift.

In order to clarify the origin of the topological Chern--Simons terms, we calculated~\cite{Gorbar:2017wpi}
the electric charge and current densities in a two-band model Hamiltonian of the electron states in Weyl semimetals.
Taking into account background electromagnetic and strain-induced pseudoelectromagnetic fields, we found that in addition to the terms given by the conventional chiral kinetic
theory~\cite{Son,Stephanov:2012ki}, there are contributions containing the information about the whole Brillouin zone.
The latter coincide exactly with the Bardeen--Zumino terms introduced in relativistic quantum field theories in order to define the consistent anomaly.
The topological origin of the Chern--Simons corrections to the electric charge and current densities is demonstrated
by expressing them in terms of the winding number in the two-band Hamiltonian.

Unlike the electromagnetic response, the chiral one induced by the pseudoelectromagnetic fields is not topologically
protected~\cite{Gorbar:2017yjn}. Although we reproduced qualitatively the anomalous chiral Hall effect~\cite{Landsteiner:2013sja,Landsteiner:2016}, the corresponding conductivity is not
topological because it depends on the parameters of the two-band model and the definition of chirality.
On the other hand, for a sufficiently small electric chemical potential, which corresponds to the well-separated Fermi surfaces of the
individual Weyl nodes, the electric current induced by the magnetic field coincides almost exactly with the current of the chiral separation
effect in linearized models~\cite{Vilenkin:1980fu,Metlitski:2005pr,Newman:2005as}. The significant deviations appear only when the Fermi surfaces undergo the Lifshitz transition, i.e., when the two separate Fermi sheets merge. As a general trend,
all chiral response coefficients tend to vanish at large values of the electric chemical potential. Therefore, in a common sense, the chirality
is well defined only in a close vicinity of the Weyl nodes.

The nontrivial topology of the electron states of multi-Weyl semimetals, i.e., the Weyl semimetals with the topological charge of the Weyl nodes
greater than one, also plays a profound role in the thermoelectric transport. In particular, our calculations
revealed~\cite{Gorbar:2017vwm} a number of thermoelectric coefficients of topological origin that describe the anomalous Ettingshausen--Nernst and
thermal Hall effects in the absence of a background magnetic field and cannot be reproduced in the conventional kinetic theory.
These effects are quantified by the off-diagonal components of the heat current induced by a thermal gradient and an electric field,
respectively. We found also that all anomalous thermoelectric coefficients in multi-Weyl semimetals contain an additional
multiplication factor, which in the limit of zero temperature and electric chemical potential, is the integer topological charge of the Weyl nodes.
Interestingly, the topological contributions to the thermal current take a form that is somewhat similar to the electromagnetic Chern--Simons currents.
However, these contributions are related to
the thermally excited quasiparticles and, thus, are not true analogs of topological electromagnetic currents.

\begin{acknowledgments}
The work of E.V.G. was partially supported by the Program of Fundamental Research of the
Physics and Astronomy Division of the National Academy of Sciences of Ukraine.
The work of V.A.M. and P.O.S. was supported by the Natural Sciences and Engineering Research Council of Canada.
The work of I.A.S. was supported by the U.S. National Science Foundation under Grants PHY-1404232
and PHY-1713950.
\end{acknowledgments}

\end{document}